\documentclass[12pt]{article}

\usepackage{algorithm}
\usepackage[noend]{algpseudocode}
\usepackage{amssymb}
\usepackage{url}
\usepackage{amsfonts} 
\usepackage{amsmath} 
\usepackage{xcolor}
\newcommand*\diff{\mathop{}\!\mathrm{d}}

\usepackage{amsmath}
\usepackage{amsfonts}
\usepackage{amssymb}
\usepackage{graphicx}

\usepackage{fullpage}
\usepackage{setspace}
\usepackage{parskip}
\usepackage{titlesec}
\usepackage[section]{placeins}
\usepackage{xcolor}
\usepackage{breakcites}
\usepackage{lineno}
\usepackage{hyphenat}

\PassOptionsToPackage{hyphens}{url}
\usepackage[colorlinks = true,
            linkcolor = blue,
            urlcolor  = blue,
            citecolor = blue,
            anchorcolor = blue]{hyperref}
\usepackage{etoolbox}

\renewenvironment{abstract}
  {{\bfseries\noindent{\abstractname}\par\nobreak}\footnotesize}
  {\bigskip}

\titlespacing{\section}{0pt}{*3}{*1}
\titlespacing{\subsection}{0pt}{*2}{*0.5}
\titlespacing{\subsubsection}{0pt}{*1.5}{0pt}

\usepackage{authblk}

\usepackage{graphicx}
\usepackage[space]{grffile}
\usepackage{latexsym}
\usepackage{textcomp}
\usepackage{longtable}
\usepackage{tabulary}
\usepackage{booktabs,array,multirow}
\usepackage{amsfonts,amsmath,amssymb}
\providecommand\citet{\cite}
\providecommand\citep{\cite}

\newif\iflatexml\latexmlfalse

\AtBeginDocument{\DeclareGraphicsExtensions{.pdf,.PDF,.eps,.EPS,.png,.PNG,.tif,.TIF,.jpg,.JPG,.jpeg,.JPEG}}

\usepackage[utf8]{inputenc}
\usepackage[english]{babel}

\begin{document}

\title{Scaling advantage of nonrelaxational dynamics for high-performance combinatorial optimization}

\author[1,2]{Timoth\'{e}e Leleu}
\author[3,4]{Farad Khoyratee} 
\author[5,6]{Timoth\'{e}e Levi}
\author[7,8]{Ryan Hamerly}
\author[1,2,6]{Takashi Kohno}
\author[1,2]{Kazuyuki Aihara}

\affil[1]{Institute of Industrial Science, University of Tokyo, Japan}
\affil[2]{International Research Center for Neurointelligence, University of Tokyo, Japan}
\affil[3]{Theoretical Quantum Physics Laboratory, RIKEN Cluster for Pioneering Research, Saitama, Japan}
\affil[4]{Physics Department, The University of Michigan, Ann 
Arbor, MI, USA}
\affil[5]{IMS, University of Bordeaux, France}
\affil[6]{LIMMS/CNRS-IIS, the University of Tokyo, Japan}
\affil[7]{Massachusetts Institute of Technology, Cambridge, MA, USA}
\affil[8]{NTT Research, 1950 University Ave \#600, East Palo Alto, California 94303, USA}

\vspace{-1em}

  \date{\today}

\begingroup
\let\center\flushleft
\let\endcenter\endflushleft
\maketitle
\endgroup

\selectlanguage{english}

\begin{abstract}
The development of physical simulators, called Ising machines, that sample from low energy states of the Ising Hamiltonian has the potential to drastically transform our ability to understand and control complex systems. However, most of the physical implementations of such machines have been based on a similar concept that is closely related to relaxational dynamics such as in simulated, mean-field, chaotic, and quantum annealing. We show that nonrelaxational dynamics that is associated with broken detailed balance and positive entropy production rate can accelerate the sampling of low energy states compared to that of conventional methods. By implementing such dynamics on field programmable gate array, we show that the nonrelaxational dynamics that we propose, called chaotic amplitude control, exhibits a scaling with problem size of the time to finding optimal solutions and its variance that is significantly smaller than that of relaxational schemes recently implemented on Ising machines.
\end{abstract}

\maketitle 

\section*{Introduction}

\begin{figure}[t]
\centering
\includegraphics[width=1.00\columnwidth]{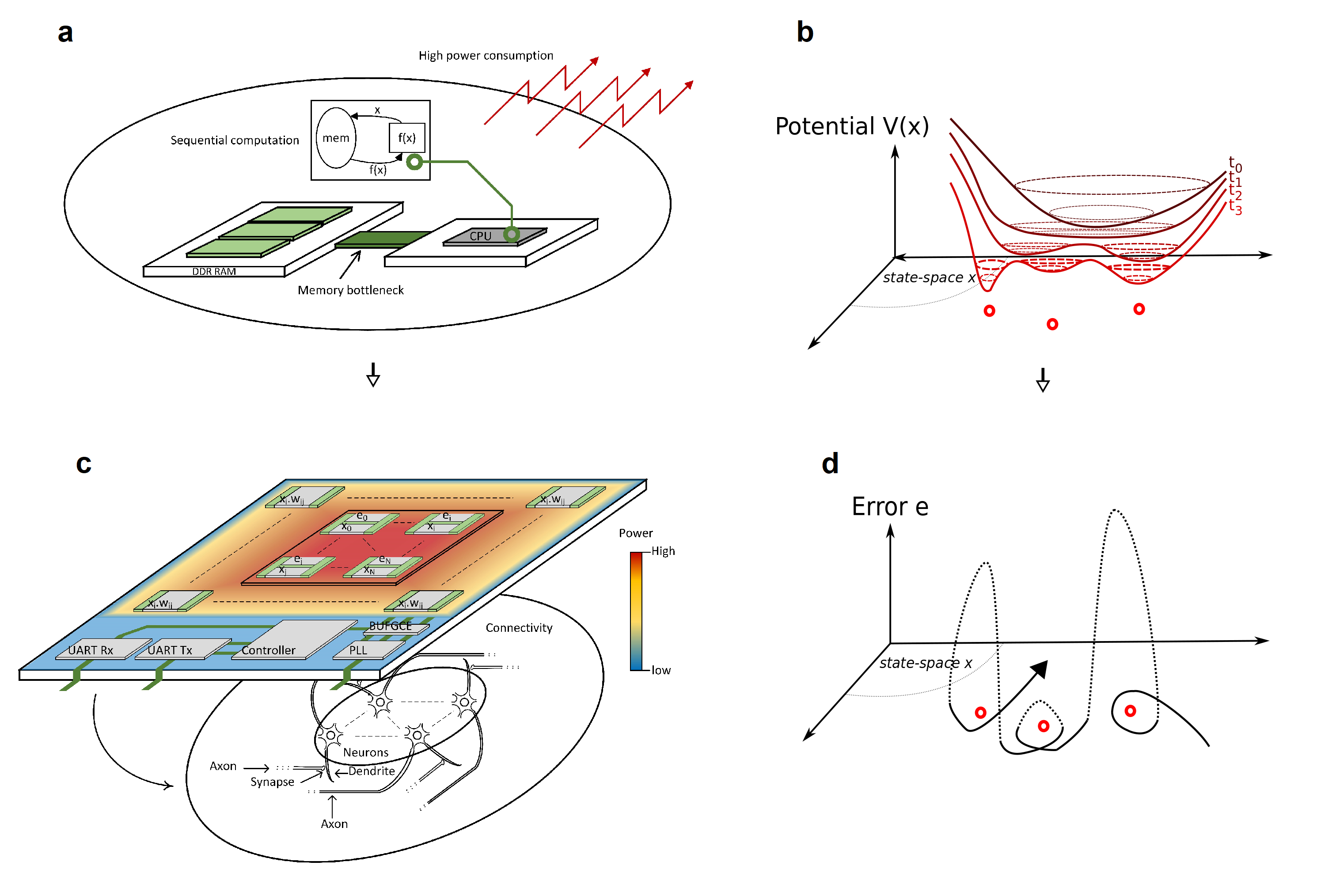}
\caption{Schematic representation of (a) conventional CPU architecture with the von Neumann bottleneck problem and (c) the proposed neuromorphic chip for combinatorial optimization. Schema of state-space dynamics of algorithms based on (b) annealing on a potential function and (d) the proposed chaotic amplitude control scheme.}
\label{fig:1}
\end{figure}

Many complex systems such as spin glasses, interacting proteins, large scale hardware, and financial portfolios, can be described as ensembles of disordered elements that have competing frustrated interactions\cite{Mezard1986} and rugged energy landscapes. There has been a growing interest in using physical simulators called ``Ising machines'' in order to reduce time and resources needed to identify configurations that minimize their total interaction energy, notably that of the Ising Hamiltonians $\mathcal{H}$ with $\mathcal{H}(\boldsymbol{\sigma}) = - \frac{1}{2} \sum_{ij} \omega_{ij} \sigma_i \sigma_j$ (with $\omega_{ij}$ the symmetric Ising couplings, i.e., $\omega_{ij} = \omega_{ji}$, and $\sigma_i = \pm 1$) that is related to many nondeterministic polynomial-time hard (NP-hard) combinatorial optimization problems and various real-world applications\cite{Kochenberger2014} (see \cite{Vadlamani2020} for a review). Recently proposed implementations include memresistor networks\cite{Cai2020}, micro- or nano-electromechanical systems\cite{Mahboob2016}, micro-magnets\cite{Camsari2017,Camsari2019}, coherent optical systems\cite{Marandi2014}, hybrid opto-electronic hardware\cite{Mcmahon2016,Inagaki2016,Pierangeli2019}, integrated photonics\cite{Roques2020,Prabhu2020,Okawachi2020}, flux qubits\cite{Hamerly2019b}, and Bose-Einstein condensates\cite{Kalinin2018}. In principle, these physical systems often possess unique properties, such as coherent superposition in flux qubits\cite{Johnson2011} and energy efficiency of memresistors\cite{Kumar2017,Cai2020}, which could lead to a distinctive advantage compared to conventional computers (see Fig.~\ref{fig:1} (a)) for the sampling of low energy states. In practice, the difficulty in constructing connections between constituting elements of the hardware is often the main limiting factor to scalability and performance for these systems\cite{Heim2015,Hamerly2019b}. Moreover, these devices often implement schemes that are directly related to the concept of annealing (either simulated\cite{Aramon2019,kirkpatrick1983}, mean-field\cite{King2018,Bilbro1989}, chaotic\cite{Kumar2017,Chen1995}, and quantum\cite{Kadowaki1998,Johnson2011}) in which the escape from the numerous local minima\cite{Tanaka1980} and saddle points\cite{Biroli1999} of the free energy function can only be achieved under very slow modulation of the control parameter (see Fig.~\ref{fig:1} (b)). These methods are dependent on non-equilibrium dynamics called aging that, according to recent numerical studies\cite{Bernaschi2020}, is strongly non-ergodic and seems to explore only a confined subspace determined by the initial condition rather than wander in the entire configurational space\cite{Cugliandolo1994} for mean-field spin glass models. In general, such systems find the solutions of minimal energy only after many repetitions of the relaxation process.

Interestingly, alternative dynamics that is not based on the concepts of annealing and relaxation may perform better for solving hard combinatorial optimization problems\cite{Leleu2019,Ercsey-ravasz2011,Molnar2018}. Various kinds of dynamics have been proposed\cite{Aspelmeier2019,Boettcher2001,Zarand2002,Leleu2017,Vadlamani2020}, notably chaotic dynamics\cite{Aspelmeier2006,Hasegawa1997,Kumar2017,Horio2008,Aihara2002}, but have either not been implemented onto specialized hardware\cite{Montanari2018,Aspelmeier2006} or use chaotic dynamics merely as a replacement to random fluctuations\cite{Hasegawa1997,Kumar2017}. We have recently proposed that the control of amplitude in mean-field dynamics can significantly improve the performance of Ising machines by introducing error correction terms (see Fig.~\ref{fig:1} (d)), effectively doubling the dimensionality of the system, whose role is to correct the amplitude heterogeneity\cite{Leleu2019}. Because of the similarity of such dynamics with that of a neural network, it can be implemented especially efficiently in electronic neuromorphic hardware where memory is distributed with the processing\cite{Furber2014,Davies2018,Benjamin2014}. In this paper, we show that this nonrelaxational dynamics is able to escape at a much faster rate than relaxational ones from local minima and saddles of the corresponding energy function. Importantly, the exponential scaling factor with respect to system size of the time needed to reach ground-state configurations of spin glasses, called time to solution, is significantly smaller in the former case, which raises the question whether this nonrelaxational dynamics is qualitatively different from the very slow relaxation observed in classic Monte Carlo simulations of spin glasses. In order to extend numerical analysis to large problem sizes and limit finite-size effects, we implement a version of nonrelaxational dynamics that we name chaotic amplitude control (CAC) on a field programmable gate array (FPGA, see Fig.~\ref{fig:1} (c)) and show that the developed hardware can be faster for finding ground-states in the limit of large problem sizes than many state-of-the-art algorithms and Ising machines for some reference benchmarks with enhanced energy efficiency.


\section*{Results}

For the sake of simplicity, we consider the classical limit of Ising machines for which the state space is often naturally described by analog variables (i.e., real numbers) noted $x_i$ in the following. The variables $x_i$ represent measured physical quantities such as voltage\cite{Cai2020} or optical field amplitude\cite{Marandi2014,Mcmahon2016,Inagaki2016,Pierangeli2019,Roques2020,Prabhu2020} and these systems can often be simplified to networks of interacting nonlinear elements whose time evolution can be written as follows:

\begin{align} \label{eq:x}
\frac{\diff x_i}{\diff t} = f_i(x_i) + \beta_i(t) \sum_{j} \omega_{ij} g_j(x_j) + \sigma_0 \eta_i,
\end{align}


\noindent where $f_i$ and $g_i$ represent the nonlinear gain and interaction, respectively, and are assumed to be monotonic, odd, and invertible ``sigmoidal'' functions for the sake of simplicity; $\eta_i$, experimental white noise of standard deviation $\sigma_0$ with $\langle \eta_i(t) \eta_j(t') \rangle= \delta_{ij} \delta(t-t')$\footnote{$\delta_{ij}$; $\delta(t)$, the Kronecker delta symbol and Dirac delta function, respectively.}; and $N$, the number of spins. Ordinary differential equations similar to eq.~(\ref{eq:x}) have been used in various computational models that are applied to NP-hard combinatorial optimization problems such as Hopfield-Tank neural networks\cite{Hopfield1985}, coherent Ising machines\cite{Wang2013}, and correspond to the ``soft'' spin description of frustrated spin systems\cite{Sompolinsky1982}. Moreover, the steady states of eq.~(\ref{eq:x}) correspond to the solutions of the ``naive'' Thouless-Anderson-Palmer (nTAP) equations that arise from the mean-field description of Sherrington-Kirkpatrick spin glasses when the Onsager reaction term has been discarded\cite{Thouless1977}. In the case of neural networks in particular, the variables $x_i$ and constant parameters $\omega_{ij}$ correspond to firing rates of neurons and synaptic coupling weights, respectively.



It is well known that, when $\beta_i = \beta$ for all $i$ and the noise is not taken into account ($\sigma_0=0$), the time evolution of this system is motion in the state space that seeks out minima of a potential function\cite{Hopfield1984} (or Lyapunov function)  $V$ given as $V = \beta \mathcal{H}(\boldsymbol{y}) + \sum_i V_b(y_i)$ where $V_b$ is a bistable potential with $V_b(y_i) = - \int_{0}^{y_i} f_i(g_i^{-1}(y)) \diff y$ and $\mathcal{H}(\boldsymbol{y}) = - \frac{1}{2} \sum_{ij} \omega_{ij} y_i y_j$ is the extension of the Ising Hamiltonian in the real space with $y_i = g_i(x_i)$ (see Supplementary Materials S1.1). The bifurcation parameter $\beta$, which can be interpreted as the inverse temperature of the naive TAP equations\cite{Thouless1977}, the steepness of the neuronal transfer function in Hopfield-Tank neural networks\cite{Hopfield1985}, or to the coupling strength in coherent Ising machines\cite{Marandi2014,Mcmahon2016,Inagaki2016}, is usually decreased gradually in order to improve the quality of solutions found. This procedure has been called mean-field annealing\cite{Bilbro1989}, and can be interpreted as a quasi-static deformation of the potential function $V$ (see Fig.~\ref{fig:1} (b)). There is, however, no guarantee that a sufficiently slow deformation of the landscape $V$ will ensure convergence to the lowest energy state contrarily to the quantum adiabatic theorem\cite{Farhi2000} or the convergence theorem of simulated annealing\cite{Granville1994}\footnote{At fixed $\beta$, global convergence to the minimum of the potential $V$ can be assured if $\sigma_0$ is gradually decrease with $\sigma_0(t)^2 \sim \frac{c}{log(2+t)}$ and $c$ sufficiently large\cite{Geman1986}. The parameter $\sigma_0^2$ is analogous to the temperature in simulated annealing in this case. The global minimum of the potential $V$ does not, however, generally correspond to that of the Ising Hamiltonians $\mathcal{H}$ at finite $\beta$.}. Moreover, the statistical analysis of spin glasses suggests that the potential $V$ is highly non-convex at low temperature and that simple gradient descent very unlikely reaches the global minimum of $\mathcal{H}(\boldsymbol{\sigma})$ because of the presence of exponentially numerous local minima\cite{Tanaka1980} and saddle points\cite{Biroli1999} as the size of the system increases. The slow relaxation time of Monte-Carlo simulations of spin glasses, such as when using simulated annealing, might also be explained by similar trapping dynamics during the descent of the free energy landscape obtained from the TAP equations\cite{Biroli1999}. In the following, we consider in particular the soft spin description obtained by taking $f_i(x_i) = (-1+p)x_i - x_i^3$ and $y_i = g(x_i) = x_i$, where $p$ is the gain parameter, which is the canonical model of the system described in eq.~(\ref{eq:x}) at proximity of a pitchfork bifurcation with respect to the parameter $p$. In this case, the potential function $V_b$ is given as $V_b(x_i) = (-1+p) \frac{x_i^2}{2} - \frac{x_i^4}{4}$ and eq.~(\ref{eq:x}) can be written as $\frac{\diff x_i}{\diff t} = - \frac{\partial V}{\partial x_i}$, $\forall i$.



In order to define nonrelaxational dynamics that is inclined to visit spin configurations associated with lower Ising Hamiltonian, we introduce error signals, noted $e_i \in \mathbb{R}$, that modulate the strength of coupling $\beta_i$ to the $i$th nonlinear element such that $\beta_i(t)$ defined in eq.~(\ref{eq:x}) is expressed as $\beta_i(t) = \beta e_i(t)$ with $\beta > 0$. The time evolution of the error signals $e_i$ are given as follows\cite{Leleu2019}:

\begin{equation} \label{eq:error}
\frac{\diff e_i}{\diff t} = - \xi (g(x_i)^2-a) e_i,
\end{equation}

\noindent where $a$ and $\xi$ are the target amplitude and the rate of change of error variables, respectively, with $a>0$ and $\xi>0$. If the system settles to a steady state, the values $y_i^* = g(x_i^*)$ become exactly binary with $y_i^* = \pm \sqrt{a}$. When $p<1$, the internal fields $h_i$ at the steady state, defined as $h_i = \sum_j \omega_{ij} \sigma_j$ with $\sigma_j = y_j^*/|y_j^*|$, are such that $h_i \sigma_i > 0$, $\forall i$\cite{Leleu2019}. Thus, each equilibrium point of the analog system corresponds to that of a zero-temperature local minimum of the binary spin system. 

The dynamics described by the coupled equations (\ref{eq:x}) and (\ref{eq:error}) is not derived from a potential function because error signals $e_i$ introduce asymmetric interactions between the $x_i$ and the computational principle is not related to a gradient descent. Rather, the addition of error variables results in additional dimensions in the phase space via which the dynamics can escape local minima. The mechanism of this escape can be summarized as follows. It can be shown (see the Supplementary Materials S1.2) that the dimension of the unstable manifold at equilibrium points corresponding to local minima $\boldsymbol{\sigma}$ of the Ising Hamiltonian depends on the number of eigenvalues $\mu(\boldsymbol{\sigma})$ with $\mu(\boldsymbol{\sigma}) > F(a)$ where $\mu(\boldsymbol{\sigma})$ are the eigenvalues of the matrix $\{\frac{\omega_{ij}}{|h_i|}\}_{ij}$ (with internal field $h_i$) and $F$ a function given as $F(y) = \frac{\psi'(y)}{\psi(y)}y$ and $\psi(y) = \frac{f(g^{-1}(y))}{(g^{-1})'(y)}$. Thus, there exists a value of $a$ such that all local minima (including the ground state) are unstable and for which the system exhibits chaotic dynamics that explores successively candidate boolean configurations. The energy is evaluated at each step and the best configuration visited is kept as the solution of a run. Interestingly, this chaotic search is particularly efficient for sampling configurations of the Ising Hamiltonian close to that of the ground state using a single run although the distribution of sampled states is not necessarily given by the Boltzmann distribution. Note that the use of chaotic dynamics for solving Ising problems has been discussed previously\cite{Goto2016,Kumar2017}, notably in the context of neural networks, and it has been argued that chaotic fluctuations may possess better properties than Brownian noise for escaping from local minima traps. In the case of the proposed scheme, the chaotic dynamics is not merely used as a replacement to noise. Rather, the interaction between nonlinear gain and error-correction results in the destabilization of states associated with lower Ising Hamiltonian. 

Ensuring that fixed points are locally unstable does not guarantee that the system does not relax to periodic and chaotic attractors. We have previously proposed that non-trivial attractors can also be destabilized by ensuring the positive rate of entropy production using a modulation of the target amplitude\cite{Leleu2019}. In this paper, we propose an alternative heuristic modulation of the target amplitude that is better suited for a digital implementation than the one proposed in \cite{Leleu2019}. Because the value of $a$ for which all local minima is unstable is not known \textit{a priori}, we propose instead to destabilize the local minima traps by dynamically modulating $a$ depending on the visited configurations $\boldsymbol{\sigma}$ as follows:

\begin{align} \label{eq:amod}
a(t) = \alpha - \rho \text{tanh}(\delta \Delta \mathcal{H}(t)),
\end{align}

\noindent where $\Delta \mathcal{H}(t) = \mathcal{H}_{\text{opt}}-\mathcal{H}(t)$; $\mathcal{H}(t)$, the Ising Hamiltonian of the configuration visited at time $t$; and $\mathcal{H}_{\text{opt}}$, a given target energy. In practice, we set $\mathcal{H}_{\text{opt}}$ to the lowest energy visited during the current run, i.e., $\mathcal{H}_{\text{opt}}(t) = \text{min}_{t' \leq t} \mathcal{H}(t')$. The function $\text{tanh}$ is the tangent hyperbolic. $\rho$ and $\delta$ are positive real constants. In this way, configurations that have much larger Ising energy than the lowest energy visited are destabilized more strongly due to smaller target amplitude $a$. Lastly, the parameter $\xi$ (see eq.~(\ref{eq:error})) is modulated as follows: $\frac{\diff \xi}{\diff t} = \gamma$ when $t - t_r < \Delta t$, where $t_r$ is the last time for which either the best known energy $\mathcal{H}_{\text{opt}}$ was updated or $\xi$ was reset. Otherwise, $\xi$ is reset to $0$ if $t - t_r \geq \Delta t$ and $t_r$ is set to $t$. Numerical simulations shown in the following suggest that this modulation results in the destabilization of non-trivial attractors (periodic, chaotic, etc.) for typical problem instances.




\begin{figure}[t]
\centering
\includegraphics[width=1.0\columnwidth]{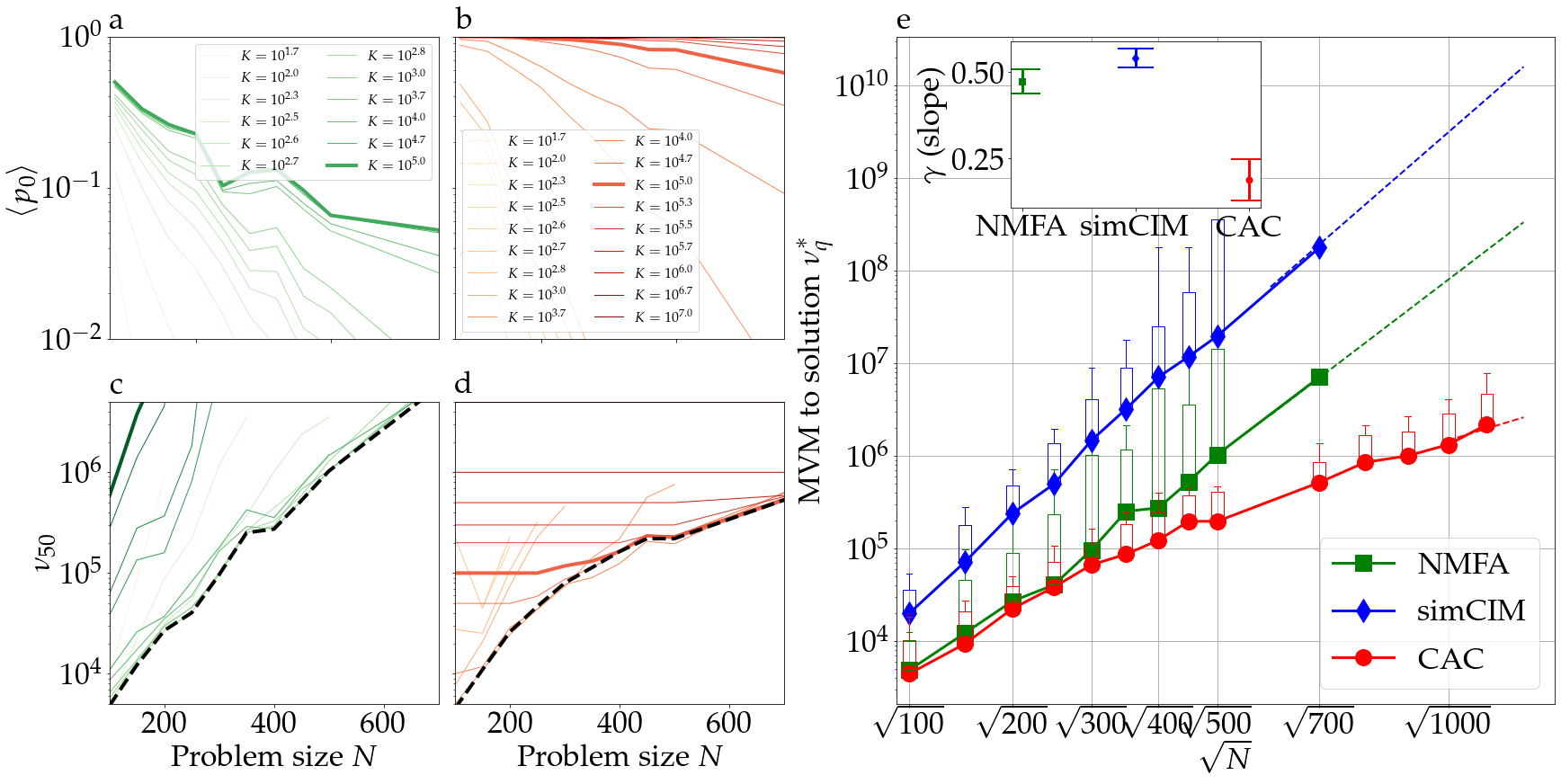}
\caption{(a,b) Average success probability $\langle p_0 \rangle$ of finding the ground state configuration of 100 Sherrington-Kirkpatrick spin glass instances and (c,d) $50^{\rm th}$ percentile of the MVM to solution distribution $\nu_{50}$ vs. system size $N$. (a,c) NMFA. (b,d) CAC. (e) Number of matrix-vector multiplication MVM to solution distribution. Lower, higher, and upper whisker of boxes show the $50{th}$, $80{th}$, and $90{th}$ percentiles of the distribution. The upper right inset shows the exponential scaling factor $\gamma$ of the $50{th}$ percentile with $\nu_{50} \sim e^{\gamma \sqrt{N}}$ for CAC, NMFA, and simCIM.}
\label{fig:2}
\end{figure}

In order to verify that the nonrelaxational dynamics of chaotic amplitude control is able to accelerate the search of mean-field dynamics for finding the ground state of typical frustrated systems, we look for the ground-states of Sherrington-Kirkpatrick (SK) spin glass instances using the numerical simulation of eqs.~(\ref{eq:x}) to (\ref{eq:amod}) and compare time to solutions with those of two closely related relaxational schemes: noisy mean-field annealing (NMFA)\cite{King2018} and the simulation of the coherent Ising machine (simCIM) \footnote{See Supplementary Materials S4.1-2}. Because the arithmetic complexity of calculating one step of these three schemes is dominated by the matrix-vector multiplication (MVM), it is sufficient for the sake of comparison to count the number of MVM, noted $\nu$, to find the ground state energy of a given instance with $99\%$ success probability, with $\nu(K) = K \frac{\text{ln}(1-0.99)}{\text{ln}(1-p_0(K))}$ and $p_0(K)$ the probability of visiting a ground state configuration at least once after a number $K$ of MVMs \textit{in a single run}. In Fig.~\ref{fig:2}, NMFA (a) and the CAC (b) are compared using the averaged success probability $\langle p_0 \rangle$ of finding the ground state for 100 randomly generated SK spin glass instances per problem size $N$. Note that the success probability of the mean-field annealing method does not seem to converge to 1 even for large annealing time (see Fig.~\ref{fig:2} (a)). Because the success probability of NMFA and simCIM remains low at larger problem size, its correct estimation requires simulating a larger number of runs which we achieved by using GPU implementations of these methods. On the other hand, the average success probability $\langle p_0 \rangle$ of CAC is of order 1 when the maximal number of MVM is large enough, suggesting that the system rarely gets trapped in local minima of the Ising Hamiltonian or non-trivial attractors. In Figure \ref{fig:2} (c) and (d) are shown the $q^{\rm th}$ percentile (with $q=50$, i.e., the median) of the MVM to solution distribution, noted $\nu_q(K;N)$, for various duration of simulation $K$, where $K$ is the number of MVMs of a single run. The minimum of these curves, noted $\nu_q^*(N)$ with $\nu_q^*(N) = \text{min}_K \nu_q(K;N)$, represents the optimal scaling of MVM to solution vs. problem size $N$\cite{Hamerly2019b}. Using the hypothesis of an exponential scaling with the square root of problem size $N$, CAC exhibits significantly smaller scaling exponent ($\gamma = 0.18 \pm 0.06$) than the NMFA ($\gamma = 0.47 \pm 0.04$) and simCIM ($\gamma = 0.54 \pm 0.03$, see inset in Fig.~\ref{fig:2} (e)). We have verified that this scaling advantage holds for various parameters of the mean-field annealing (see Supplementary Materials S4.1). Note that a root-exponential scaling of the median time to solution has been previously reported for SK spin glass problems\cite{Hamerly2019b,Patel2020} and other NP-Hard problems\cite{Hoos2014}.



Although comparison of algebraic complexity indicates that CAC has a scaling advantage over mean-field annealing, it is in practice necessary to compare its physical implementation against other state-of-the-art methods because the performance of hardware depends on other factors such as memory access and information propagation delays. To this end, CAC is implemented into a FPGA because its similarity with neural networks makes it well-fitted for a design where memory is distributed with processing (see Supplementary Materials S2 for the details of the FPGA implementation). The organization of the electronic circuit can be understood using the following analogy. Pairs of analog values $x_i$ and $e_i$, which represent averaged activity of two types of neurons, are encoded within neighboring circuits. This micro-structure is repeated $N$ times on the whole surface of the chip which resembles the columnar organization of the brain. The nonlinear processes $f_i(x_i)$, which model the local-population activation functions and are independent for $i \neq j$, are calculated in parallel. The coupling between elements $i$ and $j\in \{1,\ldots,N\}$ that is achieved by the dot product in eq.~(\ref{eq:x}) is implemented by circuits that are at the periphery of the chip and are organized in a summation tree reminiscent of dendritic branching (see Fig.~\ref{fig:1} (c)). The power consumption of the developed hardware never exceeds 5W because of limitations of the development board that we have used.



\begin{figure}
\centering
\includegraphics[width=1.0\columnwidth]{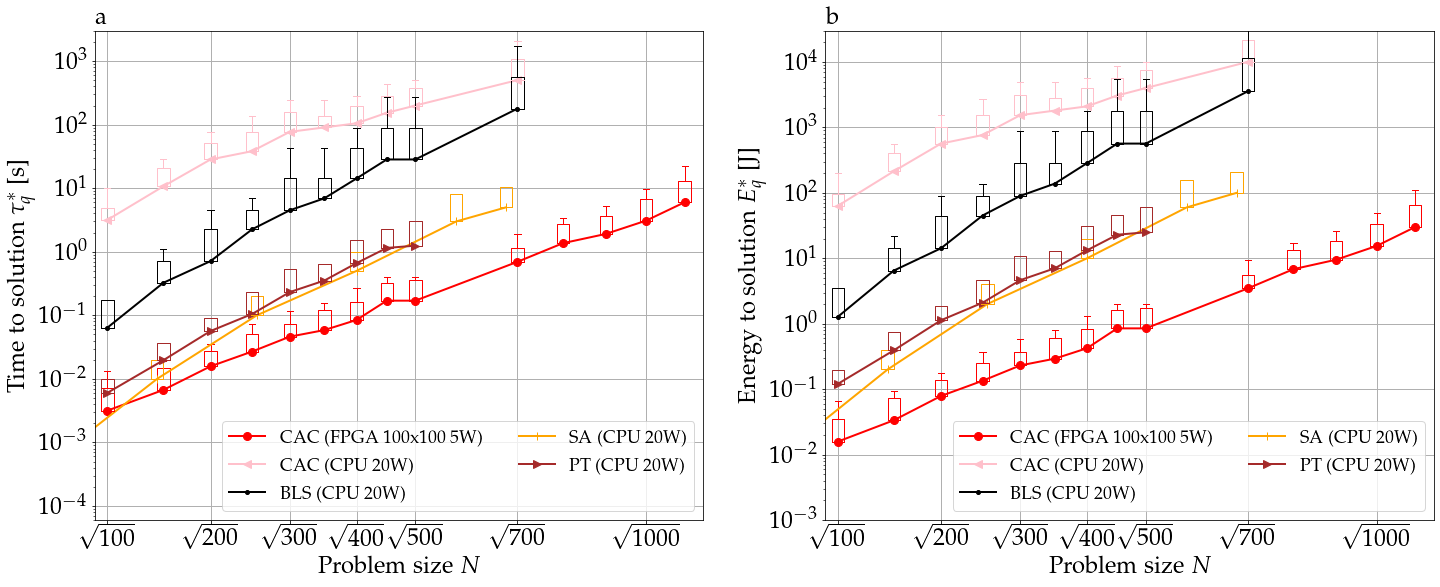}
\caption{(a) Lower, higher, and upper whisker of boxes show the $50^{\rm th}$, $80^{\rm th}$, and $90^{\rm th}$ percentiles of the time to solution distribution in seconds for the FPGA implementation of CAC with a maximum of 5W power consumption, CAC, SA, and PT algorithm running on a CPU (20W). The dashed and dotted red lines show the predictions of the real time to solution in the case of a 10W and 20W FPGA implementation, respectively. (b) The same as (a) for the energy-to-solution $E^*$.}
\label{fig:3}
\end{figure}

First, we compare the FPGA implementation of CAC against state-of-the-art CPU algorithms: break-out local search\cite{Benlic2013} (BLS) that has been used to find many of the best known maximum-cuts (equivalently, Ising energies) from the GSET benchmark set\cite{gset}, a well-optimized single-core CPU implementation of parallel tempering (or random replica exchange Monte-Carlo Markov chain sampling)\cite{Hukushima1996,Mandra2019} (PT, courtesy of S. Mandra), simulated annealing (SA)\cite{Isakov2015}. Figure \ref{fig:3} (a) shows that the CAC on FPGA has the smallest real time to solution $\tau_{q}^*$ against most other state-of-the-art algorithms despite just 5W power-consumption where $\tau_q^*(N)$ is the optimal $q^{\rm th}$ percentile of time to solution with 99\% success probability and is given as $\tau_q^*(N) = \text{min}_T \tau_q(T;N)$ where $\tau(T)$ of a given instance is $\tau(T) = T \frac{\text{ln}(1-0.99)}{\text{ln}(1-p_0(T))}$ and $T$ is the duration in seconds of a run. The probability $p_0(T)$ is evaluated using 100 runs per instance. The results of the CAC algorithm run on a CPU are also included in Fig.~\ref{fig:3}. The CPU implementation of CAC written in python for this work is not optimized and is consequently slower than other algorithms. However, its scaling of time to solution with problem size is consistent with that of CAC on FPGA. Figure \ref{fig:3} shows that CAC implemented on either CPU or FPGA has a significantly smaller increase of time to solution with problem size than SA run on CPU. Note that the power consumption of transistors in the FPGA and CPU scales proportionally to their clock frequencies. In order to compare different hardware despite the heterogeneity in their power consumption, the $q^{\rm th}$ percentile of energy-to-solution $E_q^*$, i.e., the energy $E_q^*$ required to solve SK instances with $E_q^* = P \tau_q^*$ and $P$ the power consumption\footnote{For the sake of simplicity, we assume a 20 watts power consumptions for the CPU. These numbers represent typical orders to magnitude for contemporary digital systems.}, is plotted in Fig.~\ref{fig:3} (b). CAC on FPGA is $10^2$ to $10^3$ times more energy efficient than state-of-the-art algorithms running on classical computers. 

\begin{figure}[ht]
    \centering
    \includegraphics[width=0.7\columnwidth]{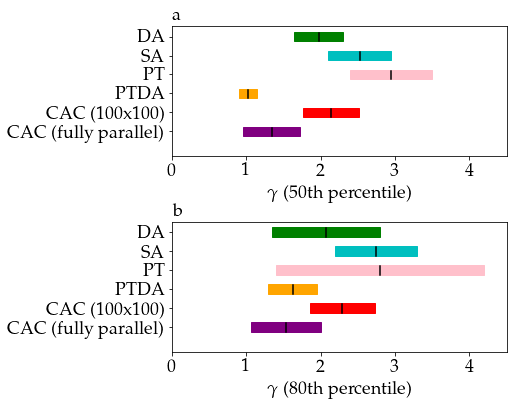}
    \caption{\label{fig:6} Scaling exponents $\frac{\gamma}{\text{log}(10)}$ of the $50^{\rm th}$ (a) and $80^{\rm th}$ (b) percentiles of the time to solution distribution based on the hypotheses of scaling in $e^{\gamma N}$ obtained by fitting data shown in Fig.~\ref{fig:2} of the proposed nonrelaxational dynamics and the scaling exponents reported in \cite{Aramon2019}. Colored boxes show the $90\%$ confidence interval in the scaling exponents. SA: simulated annealing; PT: parallel tempering; DA: digital annealer; PTDA: parallel tempering on DA. Exponents of SA, PT, DA, and PTDA are taken from \cite{Aramon2019}.}
\end{figure}

The Monte Carlo methods SA and PT have moreover been recently implemented on a special-purpose electronic chip called Digital Annealer (DA)\cite{Aramon2019}. In Fig.~\ref{fig:6}, we show the scaling exponents of 50th and 80th percentiles of the time to solution distribution for problem sizes $N=800$ to $N=1100$ based on the hypothesis of scaling in $e^{\gamma N}$ obtained by fitting data shown in Figs.~\ref{fig:2} and \ref{fig:3} (see ``CAC fully parallel'' and `CAC 100$\times$100'', respectively, in Fig.~\ref{fig:6} for the numerical values of the scaling exponents\footnote{We replicated the benchmark method of \cite{Aramon2019} for the sake of the comparison by fitting time to solution from $N=800$ up to $N=1100$.}), and compare them to that reported for SA and PT implemented on CPU and DA. The scaling obtained from fitting the time to solution in Fig.~\ref{fig:2} is based on the assumption that the matrix-vector multiplication can be calculated fully in parallel in a time that scales as $\text{log}(N)$ instead of $N^2$ (see Materials and Methods section) at least up to $N=1100$. We include this hypothesis because many other Ising machines exploit the parallelization of matrix-vector multiplication for speed up\cite{Aramon2019,Goto2021}, whereas the current implementation of CAC iterates on block matrices of size 100 by 100 and is thus only partially parallel because of resource limitations specific to the downscale FPGA used in this work. Note that the time to compute the matrix-vector multiplication is not the dominant term in the exponential scaling behavior of time to solution at large $N$. The scaling of time to solution for the nonrelaxational dynamics observed is significantly smaller than the ones of standard Monte Carlo methods SA and PT\cite{Aramon2019}, especially in the case of a fully parallel implementation. The scaling exponents of fully parallel CAC is smaller than that of DA and on par with that of PT on DA (PTDA), although CAC does not require simulating replica of the system and is thus faster in absolute time than PTDA\cite{Aramon2019}.

\begin{figure}[ht]
    \centering
    \includegraphics[width=1.0\columnwidth]{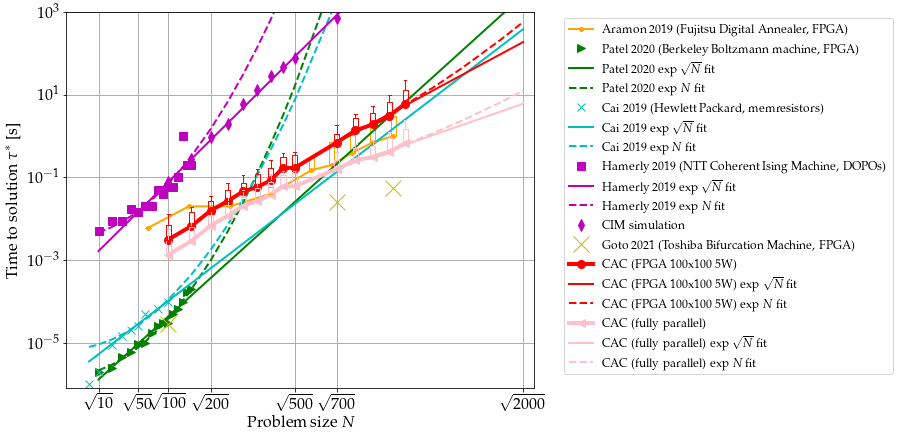}
    \caption{\label{fig:4} Median time to solution and extrapolations based on the hypotheses of scaling in $e^{\gamma N}$ and $e^{\gamma \sqrt{N}}$ by fitting the available experimental data for each Ising machine. Shaded regions show the $95\%$ confidence interval in the scaling exponents. For FPGA-CAC and DA, the lower, higher, and upper whisker of boxes show the $50^{\rm th}$, $80^{\rm th}$, and $90^{\rm th}$ percentiles of the real time to solution distribution.}
\end{figure}

Next, the proposed implementation of nonrelaxational dynamics is compared to other recently developed Ising machines (see Fig.~\ref{fig:4}). The relatively slow increase of time to solution with respect to the number of spins $N$ when solving larger SK problems using CAC suggests that our FPGA implementation is faster than the Hopfield neural network implemented using memresistors (mem-HNN)\cite{Cai2020}, the restricted Boltzmann machine using a FPGA (FPGA-RBM)\cite{Patel2020} at large $N$. Extrapolations are based on the hypotheses of scaling in $e^{\gamma N}$ and $e^{\gamma \sqrt{N}}$ by fitting the available experimental data up to $N=100$ for mem-HNN and FPGA-RBM, $N=150$ for NTT CIM, and $N=1100$ for FPGA-CAC. Figure \ref{fig:4} shows that mem-HNN, FPGA-RBM, and NTT CIM have similar scaling exponents, although FPGA-RBM tends to exhibit a scaling in $e^{\gamma N}$ rather than $e^{\gamma \sqrt{N}}$ for $N \approx 100$\cite{Patel2020}. It can be nonetheless expected that the algorithm implemented in mem-HNN, which is similar to mean-field annealing, has the same scaling behavior as simCIM and NMFA (see Fig.~2). 

It is noteworthy to mention that a recent implementation of the simulated bifurcation machine\cite{Goto2021} (SBM) which is not based on the gradient descent, similarly to CAC, but based on adiabatic evolutions of energy conservative systems performs well in solving SK problems. Both SBM and CAC exhibit smaller time to solution than other gradient based methods. SBM has been implemented on a FPGA (the Intel Stratix 10 GX) that has approximately 5 to 10 times more adaptive logic modules than the KU040 FPGA used to implement CAC. In order to compare SBM and CAC if implemented on an equivalent FPGA hardware, we plot in Fig. \ref{fig:4} the estimation of the time to solution for a fully parallel implementation of CAC using the hypothesis that one matrix-vector multiplication of size $1100\times1100$ can be achieved in $0.3\mu s$. This is the same time to compute a MVM that we can infer from time to solution reported in \cite{Goto2021} for SBM with binary connectivity given that problems of size $N=100$ ($N=700$) are solved in $29 \mu s$ ($55 ms$) and 94 (81000) MVMs, respectively. Note that SBM can reach the ground-states after approximately 20 times less MVMs than CAC at $N=100$ but only 5 times less MVM at $N=1000$, suggesting that the speed of SBM depends largely on hardware rather than an algorithmic scaling advantage. Moreover, the simulated bifurcation machine\cite{Goto2021} does not perform significantly better than our current implementation of CAC for solving instances of the reference MAXCUT benchmark set called GSET\cite{gset} (see Tab. \ref{tab:gset}) even compared to the case of the implementation on the smaller KU040 FPGA with the probability of finding maximum cuts of the GSET in a single run that is much smaller with SBM. Comparison of the scaling behavior of time to solution between CAC and SBM is unfortunately not possible based on available data\cite{Goto2021}.

\begin{table}[ht]
\tiny
\centering
\begin{tabular}{ | c | c |  c |  c |  c | c | c | c | c | c | c |}
\hline
\tiny id & N & $C^{\text{opt}}$ & $C^{\text{CAC}}$ & $C^{\text{SBM}}$ & $C^{\text{CAC}}-C^{\text{SBM}}$ & $p^{\text{CAC}}_0$ & $p^{\text{SBM}}_0$ & $<t^{\text{BLS}}>$ (s) & $<t^{\text{CAC}}>$ (s) & $<t^{\text{SBM}}>$ (s) \\ \hline
22 & 2000 & 13359 & 13359 & 13359 & 0 & 0.15 & 0.018 & 560.00 & 2.76 & \textbf{2.7} \\ 
23 & 2000 & 13344 & 13342 & 13342 & 0 & 0.25 & 0.0490 & 278.00 & (4.33) & (\textbf{0.94})\\ 
24 & 2000 & 13337 & 13337 & 13337 & 0 & 0.35 & 0.0077 & 311.00 & \textbf{2.48} & 3.1 \\ 
25 & 2000 & 13340 & 13340 & 13340 & 0 & 0.25 & 0.0022 & 148.00 & \textbf{9.49} & 15 \\ 
26 & 2000 & 13328 & 13328 & 13328 & 0 & 0.25 & 0.0050 & 429.00 & 7.14 & \textbf{3.2} \\ 
27 & 2000 & 3341 & 3341 & 3341 & 0 & 1.00 & 0.0479 & 449.00 & 9.96 & \textbf{0.26} \\ 
28 & 2000 & 3298 & 3298 & 3298 & 0 & 0.20 & 0.0690 & 432.00 & 13.86 & \textbf{0.45} \\ 
29 & 2000 & 3405 & 3405 & 3405 & 0 & 0.20 & 0.0039 & 17.00 & 9.90 & \textbf{1.2} \\ 
30 & 2000 & 3413 & 3413 & 3413 & 0 & 0.05 & 0.0045 & 283.00 & 3.12 & \textbf{2.8} \\ 
31 & 2000 & 3310 & 3309 & 3310 & -1 & 0.25 & 0.0012 & 285.00 & (14.22) & 5.3 \\ 
32 & 2000 & 1410 & 1410 & 1410 & 0 & 0.05 & 0.0012 & 336.00 & 35.90 & \textbf{33}\\ 
33 & 2000 & 1382 & 1380 & 1382 & -2 & 0.05 & 0.0002 & 402.00 & (5.68) & 120 \\ 
34 & 2000 & 1384 & 1384 & 1384 & 0 & 0.05 & 0.0015 & 170.00 & \textbf{1.55} & 27 \\ 
35 & 2000 & 7687 & 7685 & 7685 & 0 & 0.15 & 0.0002 & 442.00 & (\textbf{12.94}) & (479) \\ 
36 & 2000 & 7680 & 7679 & 7677 & 2 & 0.20 & 0.0001 & 604.00 & (19.71) & (1597) \\ 
37 & 2000 & 7691 & 7690 & 7691 & -1 & 0.05 & 0.0001 & 444.00 & (33.87) & 1278 \\ 
38 & 2000 & 7688 & 7688 & 7688 & 0 & 0.05 & 0.0003 & 461.00 & \textbf{1.94} & 213 \\ 
39 & 2000 & 2408 & 2408 & 2408 & 0 & 0.45 & 0.0006 & 251.00 & \textbf{15.01} & 266 \\ 
40 & 2000 & 2400 & 2398 & 2400 & -2 & 0.05 & 0.0001 & 431.00 & (2.45) & 48 \\ 
41 & 2000 & 2405 & 2405 & 2405 & 0 & 0.05 & 0.0020 & 73.00 & \textbf{5.57} & 48 \\ 
42 & 2000 & 2481 & 2481 & 2479 & 2 & 0.50 & 0.0002 & 183.00 & 19.15 & (240) \\ 
\hline
\end{tabular}
\caption{\label{tab:gset} \small Performance of the FPGA implementation of CAC in finding the maximum cuts known, i.e., lowest Ising Hamiltonian known, of graphs in the GSET benchmark. $id$, $C^{\text{opt}}$, $C^{\text{CAC}}$, $C^{\text{SBM}}$ are the name of instances, best maximum cuts known from \cite{Ma2017}, the proposed method after 20 runs, and Toshiba bifurcation machine on FPGA\cite{Goto2021} (FPGA-SBM), respectively. $C^{\text{SBM}}$ is evaluated using more than 20 runs. $p^{\text{CAC}}_0$ and $p^{\text{SBM}}_0$ are the probability that FPGA-CAC and FPGA-SBM find the cut $C^{\text{CAC}}$ and $C^{\text{SBM}}$ in a single run, respectively. Moreover, $<t^{\text{BLS}}>$, $<t^{\text{CAC}}>$, and $<t^{\text{SBM}}>$ are the averaged time to solution using BLS written C++ and running on a Xeon E5440 2.83 GHz\cite{Benlic2013}, the proposed scheme implemented on the KU040 FPGA, and FPGA-SBM, respectively.}
\end{table}

\begin{figure}[t]
\centering
\includegraphics[width=1.0\columnwidth]{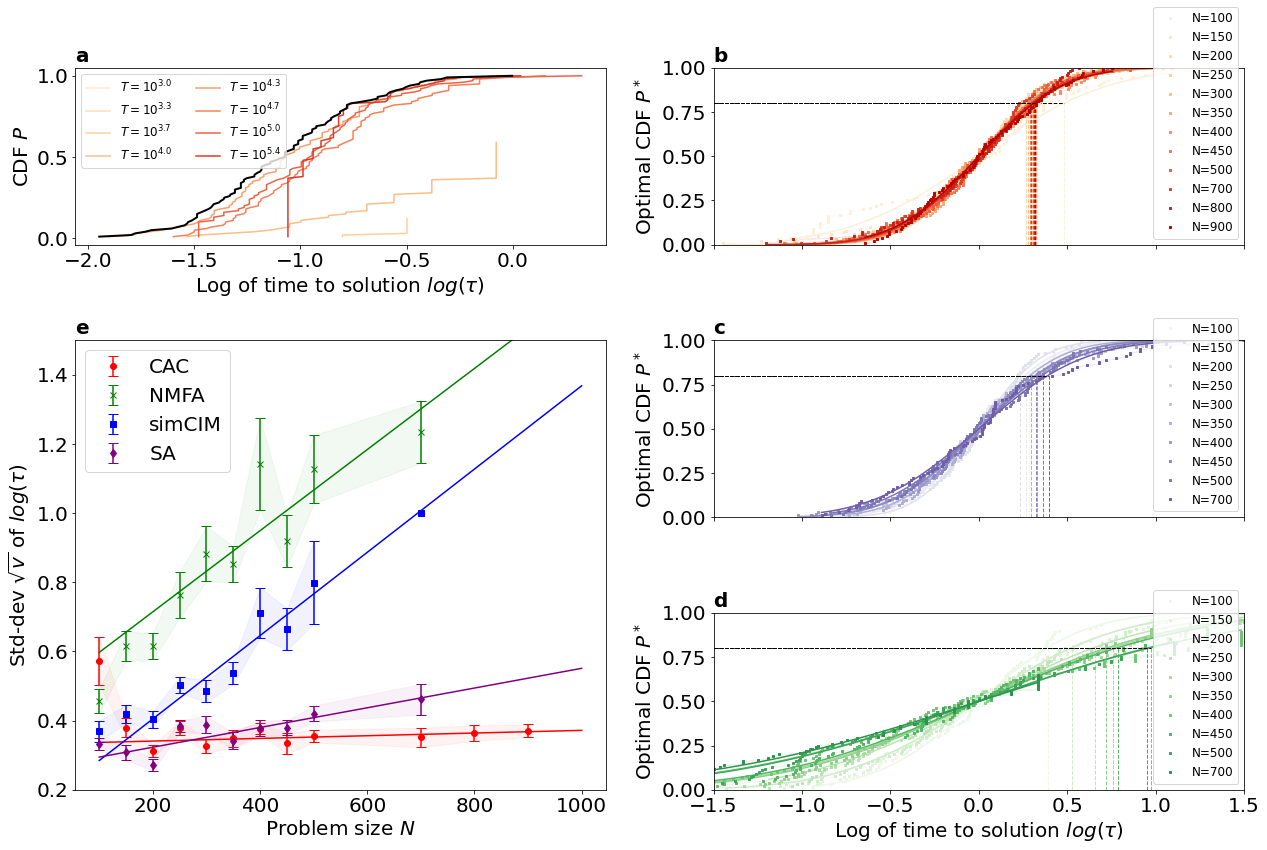}
\caption{(a) Cumulative distribution of the time to solution $P(\tau)$ for $N=400$ SK problems. (b,c,d) Optimal cumulative distribution $P^*(\tau)$ with $P^*(log(\tau)) \sim \mathcal{N}(\mu(N),\sqrt{v}(N))$ for CAC (b), SA (c), and NMFA (d), respectively. (e) Standard deviation $\sqrt{v}$ of the logarithm of time to solution distribution vs. problem size $N$. Shaded regions show the $99\%$ confidence interval in the standard deviation.}
\label{fig:5}
\end{figure}

Lastly, we consider the whole distribution of time to solution in order to compare the ability of various methods to solve harder instances. As shown in Fig.~\ref{fig:5} (a), the cumulative distribution function (CDF) $P(\tau;T)$ of time to solution with $99\%$ success probability $\tau$ is not uniquely defined as it depends on the duration $T$ of the runs. We can define an optimal CDF $P^*(\tau)$ that is independent of the runtime $T$ as follows: $P^*(\tau) = \text{max}_T P(\tau;T)$. Numerical simulations show that this optimal CDF is well described by lognormal distribution, that is $P^*(log(\tau)) \sim \mathcal{N}(\mu,\sqrt{v})$ where $\sqrt{v}$ is the standard deviation of $log(\tau)$ (see Figs. \ref{fig:5} (b), (c), and (d) for the cases of CAC, SA, and NMFA, respectively). In Fig.~\ref{fig:5} (e), it is shown that the scaling of the standard deviation $\sqrt{v}(N)$ with the problem size $N$ is significantly smaller for CAC, which implies that harder instances can be solved relatively more rapidly than using other methods. This result confirms the advantageous scaling of higher percentiles for CAC that was observed in Figs. \ref{fig:2} and \ref{fig:3}.

\section*{Conclusions}

The framework described in this paper can be extended to solve other types of constrained combinatorial optimization problems such as the traveling salesman\cite{Hopfield1985}, vehicle routing, and lead optimization problems. Moreover, it can be easily adapted to a variety of recently proposed Ising machines\cite{Cai2020,Mahboob2016,Camsari2017,Camsari2019,Marandi2014,Mcmahon2016,Inagaki2016,Pierangeli2019,Roques2020,Prabhu2020,Hamerly2019b} which would benefit from implementing a scheme that does not rely solely on the descent of a potential function. In particular, the performance of CIM\cite{Mcmahon2016,Inagaki2016}, mem-HNN\cite{Cai2020}, and chip-scale photonic Ising machine\cite{Okawachi2020}, which have small time to solution for small problem sizes\footnote{Relaxation dynamics may be faster for solving small problem sizes for which it may be sufficient to do rapid sampling based on convex optimization\cite{Ma2019}.} ($N \approx 100$) but with a relatively large scaling exponent that limit their scalability, could be significantly improved by adapting the implementation we propose if these hardware can be shown to be able to simulate larger numbers of spins experimentally. Rapid progress in the growing field of Ising machines may allow to verify scaling behaviors of the various methods at larger problem sizes and, thus, limit further finite-size effects.

The scaling exponents we have reported in this paper are based on the integration of the chaotic amplitude control dynamics using a Euler approximation of its ODEs. We have noted that the scaling of MVMs to solution of SK spin glass problems is reduced when the Euler time step is decreased (see Supplementary Materials S2.8). The scaling exponents of CAC might thus be smaller than reported in this paper in the limit of a more accurate numerical integration over continuous time. It is therefore important future work to evaluate the scaling for $N \gg 1000$ using a faster numerical simulation method. Such numerical calculations require a careful analysis of the integration method of ODEs, numerical precision, and tuning of parameters. It is also of considerable interest to implement CAC on an analog physical system for further reduction of power consumption.

Nonrelaxational dynamics described herein is not limited to artificial simulators and likely also emerge in natural complex systems. In particular, it has been hypothesized that the brain operates out of equilibrium at large scales and produces more entropy when performing physically and cognitively demanding tasks\cite{Lynn2020b,Wolf2018} in particular for the ones involving creativity for maximizing reward rather than memory retrieval. Such neural processes cannot be explained simply by the relaxation to fixed point attractors\cite{Hopfield1984}, periodic limit cycles\cite{Yan2013}, or even low-dimensional chaotic attractors whose self-similarity may not be equivalent to
complexity but set the conditions for its emergence\cite{Ageev2018,Wolf2018}. Similarly, evolutionary dynamics that is characterized as non-ergodic when the time required for representative genome exploration is longer than available evolutionary time\cite{Mcleish2015} may benefit from nonrelaxational dynamics rather than slow glassy relaxation for faster identification of high-fitness solutions. A detailed analytic comparison between the slow relaxation dynamics observed in Monte Carlo simulations of spin glasses with the one proposed in this paper is needed in order to explain the apparent difference in their scaling of time to solution exhibited by our numerical results.

\section*{Materials and Methods}

We target the implementation of a low-power system with maximum power supply of 5W using a XCKU040 Kintex Ultrascale Xilinx FPGA integrated on an Avnet board. The implemented circuit can process Ising problems of up to at least 1100 spins fully connected and of more than 2000 spins sparsely connected within the 5W power supply limit. Data are encoded into 18 bits fixed point vectors with 1 sign, 5 integer and 12 decimal bits to optimize computation time and power consumption. An important feature of our FPGA implementation of CAC is the use of several clock frequencies to concentrate the electrical power on the circuits that are the bottleneck of computation and require high speed clock. For the realization of the matrix-vector multiplication, each element of the matrix is encoded with 2 bits precision ($w_{ij}$ is $-1$, 0 or 1). An approximation based on the combination of logic equations describing the behavior of a multiplexer allows to achieve $10^4$ multiplications within one clock cycle. The results of these multiplications are summed using cascading DSP and CARRY8 connected in a tree structure. Using pipelining, a matrix-vector multiplication for a squared matrix of size $N$ is computed in $2+5\frac{\text{log}(N-4)}{\text{log}(5)}+(\frac{N}{u})^2$ clock cycles (see Supplementary Materials S2.4) at a clock frequency of 50MHz with $u=100$ which is determined by the limitation of the number of available electronic component of the XCKU040 FPGA. The block size $u$ can be made at least 3 times larger using commercially available FPGAs, which implies that the number of clock cycles needed to compute a dot product can scale almost logarithmically for problems of size $N \approx 1000$ (see Supplementary Materials S2.4 for discussions of scalability) and that the calculation time can be further significantly decreased using a higher-end FPGA. The calculation of the nonlinearity $f_i$ and error terms is achieved at higher frequency (300MHz and 100Mhz) using DSP in $8 + (N/u)$ and $9 + (N/u)$ clock cycles, respectively. In order to minimize energy resources and maximize speed, the nonlinear and error terms are calculated multiple times during the calculation of a single matrix-vector multiplication (see Supplementary Materials S2).

Prior to computing the benchmark on the Sherrington-Kirkpatrick instances, the parameters of the system (see Supplementary Materials S2.8) are optimized automatically using Bayesian optimization and bandit-based methods\cite{Falkner2018}. The automatic tuning of parameters for some previously unseen instances is out of the scope of this work but can be achieved to some extent using machine learning techniques\cite{Dunning2018}. Sherrington-Kirkpatrick instances used in this paper are available upon request. The GSET instances are available at https://web.stanford.edu/~yyye/yyye/Gset/.

\section*{Acknowledgments}

The authors thank Salvotare Mandra for providing results of the PT algorithm. This research is partially supported by the Brain-Morphic AI to Resolve Social Issues (BMAI) project (NEC corporation), AMED under Grant Number JP20dm0307009, UTokyo Center for Integrative Science of Human Behavior(CiSHuB), and Japan Science and Technology Agency Moonshot R\&D Grant Number JPMJMS2021



\selectlanguage{english}
\FloatBarrier

\pagebreak
\begin{center}
\textbf{\large Supplementary materials: Scaling advantage of nonrelaxational dynamics for high-performance combinatorial optimization}
\end{center}
\setcounter{equation}{0}
\setcounter{figure}{0}
\setcounter{table}{0}
\setcounter{page}{1}
\makeatletter
\renewcommand{\theequation}{S\arabic{equation}}
\renewcommand{\thefigure}{S\arabic{figure}}

\section{Theoretical analysis}

\subsection{Derivation of a potential function}

First, we analyze the system described by eq.~(1) only, when $\sigma_0 = 0$ and $\beta_i = \beta$, $\forall i$. In this case, the potential function $V(\boldsymbol{y}) = - \frac{1}{2} \beta \sum_{ij} \omega_{ij} y_i y_j - \sum_i \int_{0}^{y_i} f_i(g_i^{-1}(y)) \diff y$ has the following property\cite{Hopfield1984}:

\begin{align} 
\frac{\diff V}{\diff t} &= - \sum_i \frac{\diff y_i}{ \diff t} (\beta  \sum_{j} \omega_{ij} y_j + f_i(g_i^{-1}(y_i))),\\
& = - \sum_i \frac{\diff y_i}{ \diff t} (\frac{\diff x_i}{ \diff t}),\\
& = - \sum_i (\frac{\diff y_i}{ \diff t})^2 (g^{-1})'(y_i).
\end{align} 

\noindent Consequently, $V$ is such that $\frac{\diff V}{\diff t} < 0$ because $g$ is strictly monotonic and $\frac{\diff V}{\diff t} = 0 \implies \frac{\diff y_i}{\diff t} = 0 $, $\forall i$. In other words, the dynamics of the system can be understood in this case as the gradient descent on the potential function $V$ and its stable steady states correspond to local minima of $V$.

\subsection{Analysis of CAC}

Next, we analyze the system described in eqs.~(1) and (2) ($a$ is constant) by considering the change of variable $y_i = g(x_i)$ with $g$ such that $g^{-1}(y_i)=x_i$. In this case, eqs.~(1) and (2) can be rewritten as follows (note that $f$ and $g$ are odd functions):

\begin{align} 
\phi'(y_i) \frac{\diff y_i}{\diff t} &= f(g^{-1}(y_i)) + \beta e_i \sum_j \omega_{ij} y_j,\label{eq:y}\\
\frac{\diff e_i}{\diff t} &= \xi (y_i^2 - a) e_i, \label{eq:ye}
\end{align}

\noindent where $\phi(y) = g^{-1}(y)$.

We analyze the dimension of the unstable manifold of CAC by linearizing the dynamics near the steady states and calculating the real part of eigenvalues of the corresponding Jacobian matrices. The steady state of the system in eqs.~(\ref{eq:y}) and (\ref{eq:ye}) can be written as follows:

\begin{align} 
y_i^* &= \sigma_i \sqrt{a} \label{eq:std_y},\\
e_i^* &= - \frac{f(g^{-1}(\sigma_i \sqrt{a}))}{\beta \sqrt{a} h_i} =- \frac{f(g^{-1}(\sqrt{a}))}{\beta \sqrt{a} \sigma_i h_i}. \label{eq:std_e}
\end{align}

\noindent with $h_i = \sum_j \omega_{ij} \sigma_j$. The last equality is a consequence of the fact that $g$, and thus also $g^{-1}$, are odd functions. 

The Jacobian matrix $J$ corresponding to the system of eqs.~(1) and (2) at the steady state can be written in the block representation $J = [J^{yy} J^{ye}; J^{ey} J^{ee}]$ with its components given as follows:

\begin{align} 
J^{yy}_{ij} &= \psi'(\sqrt{a}) \delta_{ij} - \frac{\psi(\sqrt{a})}{\sqrt{a}} \frac{\omega_{ij}}{h_i \sigma_i},\\
J^{ye}_{ij} &= \frac{\beta \sqrt{a}}{\phi'(\sqrt{a})} h_i \delta_{ij},\\
J^{ey}_{ij} &= \frac{-2 \xi f(g^{-1}(\sqrt{a}))}{\beta h_i} \delta_{ij},\\
J^{ee}_{ij} &= 0.
\end{align}

\noindent with $\psi(y) = \frac{f(g^{-1}(y))}{\phi'(y)}$ and $\phi(y) = g^{-1}(y)$. Note that we define $J^{ey}_{ii} J^{ye}_{jj} = b$ with $b= -2 \xi \sqrt{a} \psi(\sqrt{a})$.

The eigenvalues of the Jacobian matrix J are solutions of the polynomial equation $P(\lambda) = \text{det}[J-\lambda I] = \text{det}[(J^{yy}-\lambda I)\lambda I - b I]$. The eigenvalues of $J$ are thus solutions of the quadratic equation $z(z-\lambda_i)-b = 0$ where $\lambda_i$ is the $i^{\rm th}$ eigenvalue of the matrix $J^{yy}$. Therefore, the eigenvalues of $J$ can be described by pairs $\lambda_i^+$ and $\lambda_i^-$ given as follows:

\begin{align} 
\lambda_i^{\pm} &= \frac{1}{2} (\lambda_i \pm \sqrt{\Delta_i}), \text{ with} \\
\lambda_i &= \psi'(\sqrt{a}) - \frac{\psi(\sqrt{a})}{\sqrt{a}} \mu_i, \text{ and}\\
\Delta_i &= \lambda_i^2 + 4 b,
\end{align}

\noindent where $\mu_i$ is the $i^{\rm th}$ eigenvalue of the matrix $D[(\boldsymbol{\sigma} \boldsymbol{h})^{-1}] \Omega$.

The eigenvalues $\lambda_i^{+}$ and $\lambda_i^{-}$ become complex conjugate when the $\Delta_i = 0$, i.e., $[\psi'(\sqrt{a})-\frac{\psi(\sqrt{a})}{\sqrt{a}} \mu_i]^2 - 8 \xi \sqrt{a} \psi(\sqrt{a}))=0$, which can be rewritten as follows:

\begin{align} 
\mu_i = G_{\xi}^{\pm} (\sqrt{a}),
\end{align}

\noindent with $G_{\xi}^{\pm} (y) = \frac{y}{\psi(y)}[\psi'(y) \pm 2 \sqrt{2 \xi |y| \psi(|y|) }]$.

\begin{figure}[t]
\begin{center}
\includegraphics[width=1.00\columnwidth]{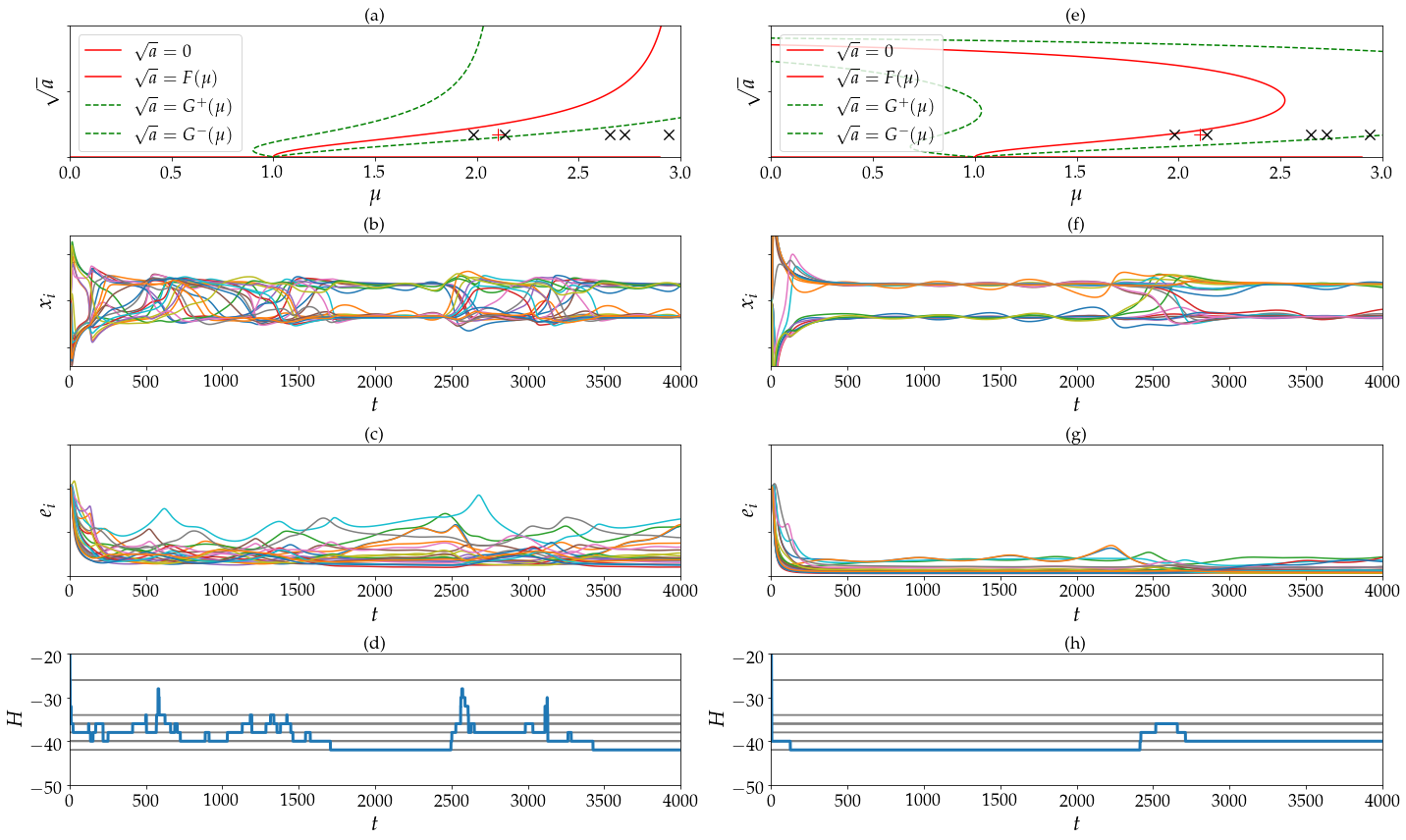}
\caption{(a,e) Bifurcation diagram in the space $\{\sqrt{a},\mu = \mu_j(\boldsymbol{\sigma})\}$, $\forall j \in \{1, \dots, N\}$, at configuration $\boldsymbol{\sigma}$. The full red and green dotted lines correspond to the set for which $Re[\lambda_i^{\pm}] = 0$ (where fixed points corresponding to local minima become stable) and $\Delta_i = 0$  (where oscillations start to occur around fixed points). Red $+$ and black $\times$ symbols correspond to the largest eigenvalues $\mu_0(\boldsymbol{\sigma})$ of the Jacobian at the ground state and excited states, respectively, of an Ising problem of size $N=15$ and $a=0.03$. In (b,f) and (c,g) are shown the dynamics of the variables $\boldsymbol{x}$ and $\boldsymbol{e}$, respectively. In (d,h) are shown the Ising Hamiltonian $\mathcal{H}(t)$ with horizontal lines showing the values of the Ising Hamiltonian at local minima. (a,b,c,d) $f(x) = (-1+p) x - x^3$ and $g(x)=x$ with $p=0.95$, $\beta = 0.1$, $\xi=0.1$. (e,f,g,h) $f(x) = -x+\text{tanh}[0.99 x]$ and $g(x)=\text{tanh}[x]$ with $\beta = 0.1$, $\xi=0.1$.}
\label{fig:supp_ana1}
\end{center}
\end{figure}

Lastly, the real part of eigenvalues $\lambda_i^{+}$ become equal to zero at the condition given as follows (note that $\text{Re}[\lambda_i^+] \geq \text{Re}[\lambda_i^-]$ and $\mu_i$ are real at local minima\footnote{Because the vector $\boldsymbol{\sigma} \cdot \boldsymbol{h}$ has positive components at local minima, the eigenvalues of $D[(\boldsymbol{\sigma} \cdot \boldsymbol{h})^{-1}] \Omega$ are the same as the ones of $D[(\boldsymbol{\sigma} \cdot \boldsymbol{h})^{-1}]^{\frac{1}{2}} \Omega D[(\boldsymbol{\sigma} \cdot \boldsymbol{h})^{-1}]^{\frac{1}{2}}$ (Sylvester's law of inertia), which is a symmetric real matrix. Thus, the eigenvalues $\mu_j$ are always real.} for which, $\forall j$, $\sigma_j h_j >0$):

\begin{align} 
\sqrt{a} & =0 \text{ or } \psi(\sqrt{a}) = 0 \text{ if } \Delta_i \leq 0,\\
\mu_i &= F(\sqrt{a}) \text{ otherwise.}
\label{eq:F}
\end{align}

\noindent with $F(y) = \frac{\psi'(y)}{\psi(y)}y$. The dimension of the unstable manifold at a given fixed point is then given by the number of indices $i$ for which $\text{Re}[\lambda_i^{\pm}]$ is positive. To illustrate the destabilization of the ground state configuration of an Ising problem, we show in Fig. \ref{fig:supp_ana1} the dynamics of CAC when solving a problem of size $N=15$ spins when the state encoding for the ground state configuration unstable for two different examples of functions $f$ and $g$ defining eqs.~(1) and (2). The first set of functions $f$ and $g$ with $f(x) = (-1+p) x - x^3$ and $g(x)=x$ shown in Fig.~\ref{fig:supp_ana1} (a,b,c,d) corresponds to the soft spin model (or simulation of CIM) with chaotic amplitude control whereas the second one (e,f,g,h) with $f(x) = -x+\text{tanh}[0.99 x]$ and $g(x)=\text{tanh}[x]$ corresponds to an Hopfield neural network with amplitude heterogeneity error correction.  These two figures show that the stability of a given local minima of the Ising Hamiltonian depends on the value of the target amplitude $a$ as predicted in eq.~(\ref{eq:F}).

\clearpage

\section{FPGA implementation}
\subsection{FPGA implementation of CAC}

CAC has been implemented on a XCKU040 Xilinx FPGA integrated into the KU040 development board designed by Avnet. The circuit can process Ising problems of more than 2000 spins. The reconfigurable chip is a regular Ultrascale FPGA including 484,800 flip-flops, 242,400 Look Up Table (LUT), 12,000 Digital Signal Processing (DSP) and 600 Block RAM (BRAM). Initial value for $x_i$, $e_i$ and $\omega_{ij}$ are sent through Universal Asynchronous Receiver Transmitter (UART) transmission. UART protocol has been chosen because it does not require a lot of resources to be implemented and its simplicity. 

Data are encoded into 18 bits fixed point with 1 sign bit, 5 integer bits which represents a good compromise between accuracy and power consumption. The power consumption of the FPGA is equal or lower to 5W depending on the problem size. The system is defined by its top-level entity that represents the highest level of organization of the reconfigurable chip.

\subsection{Top level entity}

\begin{figure}[ht]
    \centering
    \includegraphics[width=1.0\textwidth]{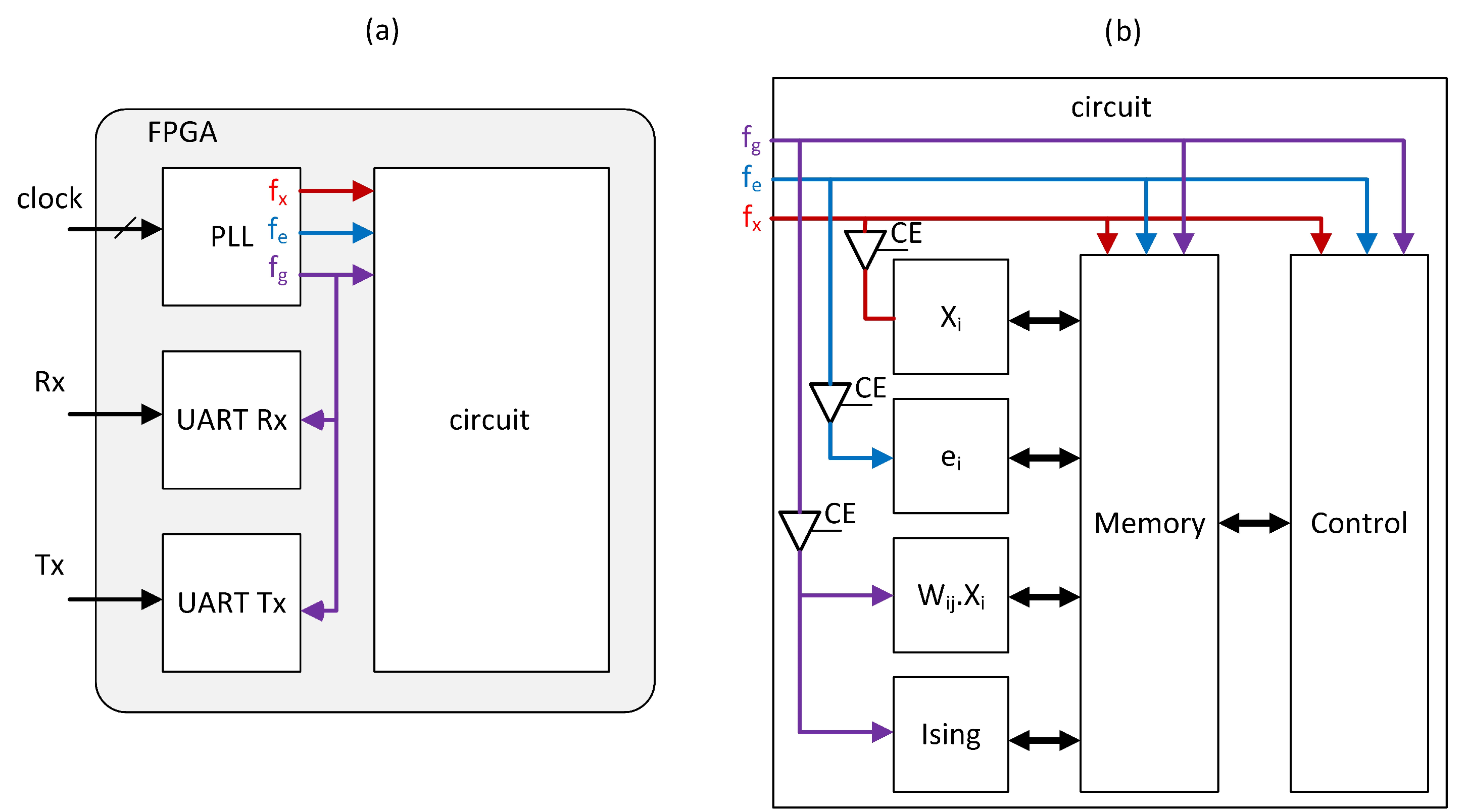}
    \caption{Organization of the FPGA circuit. (a) The top layer entity is divided into four modules: The Phased-Locked Loop (PLL) that convert a differential clock of 250MHz into three clocks $f_g$=50MHz, $f_x$=300Mhz and $f_e$=100MHz respectively representing the global clock, the clock for $x_i$ and the clock for $e_i$. Two UART modules are used to receives parameters and to return the results of the computation. (b) The Max-Cut circuit is composed of several processes aiming to control the exchange of data between the memory and the computation cores ($x_i$, $e_i$, $x_i\omega_{ij}$ and Ising). The three generated clocks are controlled by the Clock Enable (CE) port of the BUFGCE component of the FPGA.}
    \label{fig:figure1}
\end{figure}

The circuit is organized into four principal building blocks as shown on Fig.~\ref{fig:figure1} (a). Several clock frequencies have been generated to concentrate the electrical power on the circuits that need high speed clock when these circuits constitute a bottleneck in the current implementation. Finally, the organization of the main circuit is represented in Fig.~\ref{fig:figure1} (b). The control block is composed of Finite State Machines (FSM) and is well tuned to pilot all computation core, the Random-Access Memory (RAM), and data flowing between computation cores and RAM. All circuits are synchronized although the system utilizes multiple clock frequencies. Clock Enable (CE) ports on the BUFGCE component are used to stop the power when a circuit is not used in order to reduce further the global power dissipation. 

\clearpage

\subsection{Temporal organization of the computation}

The control circuit pilot the computation in order to calculate several steps of $x_{i}$ and $e_{i}$ per calculation of the Ising coupling. The dynamics of CAC (see eqs.~(1), (2) and (3)) is implemented based on the following pseudocode:

\renewcommand{\thealgorithm}{}

\begin{algorithm}
\caption{Pseudocode from which the FPGA implementation is adapted}
\label{pseudocode}
\begin{algorithmic}[1]

\For{$\nu \in \{0,\cdots,K\}$}\Comment{Iterate on the number of MVMs}

\State $\boldsymbol{x}' \gets \boldsymbol{x}$ \Comment{Save state}

\State $\boldsymbol{I} \gets \epsilon \boldsymbol{e} \dot (\Omega \boldsymbol{x}')$ \Comment{calculation of injection term}

\State $\boldsymbol{\sigma} \gets \frac{x'_i}{|x'_i|}$
\State $H \gets -\frac{1}{2} \boldsymbol{\sigma} \dot (\Omega \boldsymbol{\sigma})$ \Comment{Ising energy calculation}

\For{$i \in \{0,\cdots,n_x\}$} \Comment{Update nonlinear terms}
\State $\boldsymbol{\Delta}_x \gets (-1+p)\boldsymbol{x} - (\boldsymbol{x})^3 + \boldsymbol{I}$
\State $\boldsymbol{x} \gets \boldsymbol{x} + \boldsymbol{\Delta}_x dt_x$
\EndFor

\For{$i \in \{0,\cdots,n_y\}$} \Comment{Update error terms}
\State $\boldsymbol{\Delta}_e \gets -\beta ((\boldsymbol{x}')^2 - a) \boldsymbol{e}$
\State $\boldsymbol{e} \gets \boldsymbol{e} + \boldsymbol{\Delta}_e dt_e$
\EndFor

\State $\beta \gets \beta + \lambda dt$ \Comment{Update $\beta$}

\State $\Delta H \gets H-H_{\text{opt}}$ \Comment{Update $a$}
\State $a \gets \alpha + \rho \text{tanh}(\delta \Delta H)$

\If{$\nu-\nu_c>\tau/dt$} \Comment{Reset of $\beta$}
\State $\nu_c \gets \nu$
\State $\beta \gets 0$
\EndIf

\If{$H<H_{\text{opt}}$} \Comment{Update optimal $H$}
\State $H_{\text{opt}} \gets H$
\State $\nu_{\text{opt}} \gets \nu$
\State $\nu_{c} \gets \nu$
\EndIf

\EndFor

\end{algorithmic}
\end{algorithm}

\noindent Note that the order of operations described in the pseudocode are a simplification of the ones occurring on the FPGA. The CPU implementation of CAC used in this work is written in python. For the python simulation, variables $e_i$ saturate at the value $e_i^{\text{MAX}} = 32$ in order to approximate the digital encoding of the FPGA implementation.

\begin{figure}[ht]
    \centering
    \includegraphics[width=1.0\textwidth]{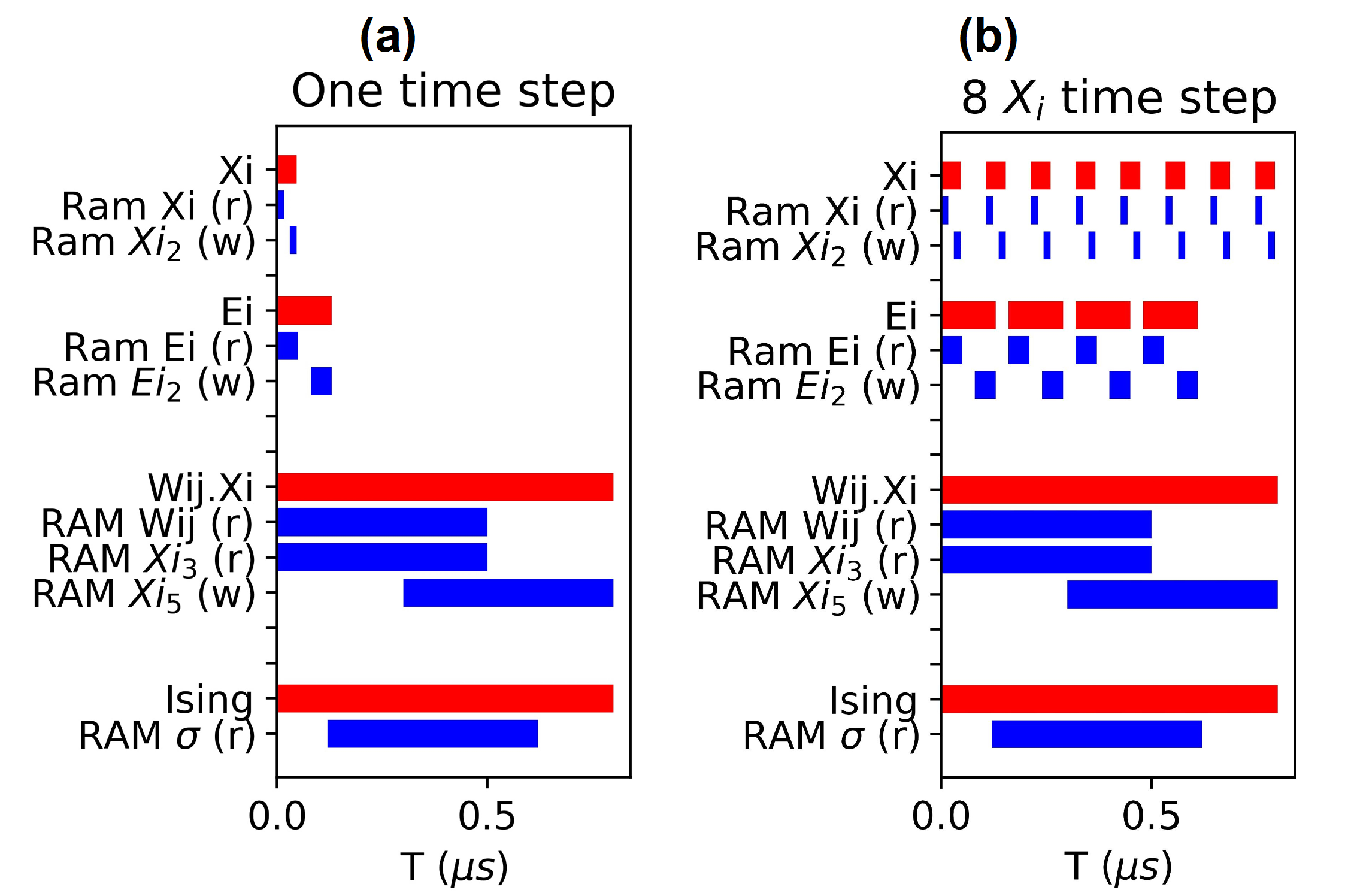}
    \caption{Temporal representation of the circuits where the red bars represent the computation cores and the blue bars the use of RAM. Here, (r) stand for read and (w) for write. (a) One time step representing the classical way of solving differential equation. In this configuration, the dot product create a bottleneck to the computation speed. (b) 8 time steps for $x_i$ and 4 time step for $x_i$ accelerating the computation time to find the ground state energy and create a new bottleneck to the number of possible time step possible. Note that the time between every red bars are necessary update the $x_i$ RAM which is not represented in this figure for better clarity.}
    \label{fig:figure2}
\end{figure}

\begin{figure}[ht]
    \centering
    \includegraphics[width=1.0\textwidth]{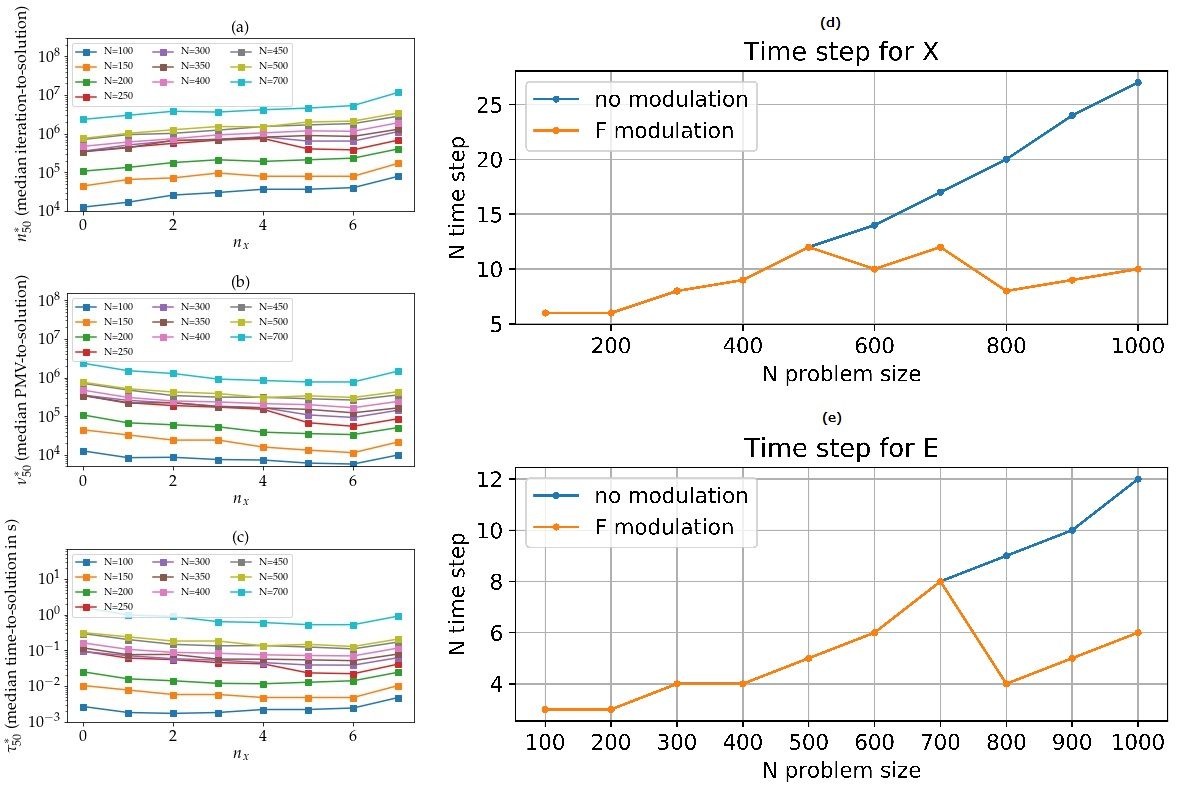}
    \caption{(a) The number of iterations of the $\boldsymbol{x}$ variable to solution vs. the number of update, noted $n_x$, of the $\boldsymbol{x}$ variable per matrix-vector product. (b) The number of matrix-vector multiplications MVM to solution. (c) Real time to solution. (d) Maximum possible time step with and without clock modulation on $x_i$. (e) Maximum possible time step with and without clock modulation on $e_i$.}
    \label{fig:figure3}
\end{figure}

The temporal organization of the circuit represented in Fig.~\ref{fig:figure2} shows how is controlled the reading and writing state on the RAM in the case of a problem size $N=500$ spins. Fig.~\ref{fig:figure2} (a) shows the case of a single $x_i$ and $e_i$ calculation per dot product which is the classic way to integrate the differential equations (1) and (2) using the Euler method. Fig.~\ref{fig:figure2} (b) represents the case of eight and four calculations of $x_i$ and $e_i$, respectively, per dot product. In this case, the error correction term is computed with a normalized time step of $2^{-4}$ (half of the time step used in the computation of $x_i$). The total power consumption can be reduced by increasing the frequency of circuits involved in the calculation of $x_i$ because these circuits are smaller than those involved in the dot product calculation (see Fig.~\ref{fig:figure3} (d) and (e)). Increasing the problem size will increase the power consumption by filling up the calculation pipelines but this can be balanced out by reducing the frequency used for the calculation of $x_i$ for larger problem size. The end of the dot product is followed by an update of all RAM and saving of $\sigma_i$ which correspond to the Most Significant Bit (MSB) of $x_i$ values. This step is not represented here.

The use of several $x_i$ and $e_i$ calculations per dot product allows to reduce the bottleneck created by the later whose calculation is in principle the most time consuming. Fig.~\ref{fig:figure3} (a) to (c) show the reduction in time to solution vs. the number of $x_i$ calculations per dot product.

\subsection{Coupling}
\subsubsection{Overview of the coupling circuit}
The dot product operation is preceded by a multiplication by $\beta$ so that the domain of $x_i$ is reduced and, in turn, the number of digits required to encode the integer part of $x_i$. The result of the dot product is multiplied by the error correction term. Since the dot product is performed at 50MHz clock speed, the optional registers of the DSP are no longer required and have been removed for the two multiplications resulting in 1 clock cycle operation for each operation.

\subsubsection{High speed and massively parallel multiplication}
The multiplication of the elements of a vector $\boldsymbol{x}$ by the element of a matrix $\Omega$ can be performed using an approximation based on a specific circuit combining logic equations describing the behavior of a multiplexer and the optimized use of FPGA components. This circuit allows the design to achieve 10,000 multiplications in 1 clock cycle. In our case an element of $\boldsymbol{x}$ is fixed point binary vector on k bits that are multiplied by an element of a matrix composed of two bits vectors where $\omega$ is an element of $\Omega$ and $\boldsymbol{\omega} \in \{-1, 0, 1\}$. The behavior of a multiplexer that implements the multiplication of $x$ by $\omega$ by selecting $R$ $\in \{-x, 0, x\}$ is given as follows ($\bar{x}$ represent $-x$):

\begin{equation}
    \label{eq:1}
    R = x \bar{\omega}_1 \bar{\omega}_0 + \bar{x} \omega_1 \omega_0
\end{equation}

\noindent where $x$ is an element of a vector, $\omega_i$ a binary vector and an element of the matrix $\Omega$ with i the index of the binary vector that is either 1, the most significant bit (MSB) or 0, the less significant bit (LSB). If we expand eq.~(\ref{eq:1}) to each bits $x_i$ of the vector $x$ and use Boolean operation, we obtain the equation eq.~(\ref{eq:2}) given as follows:

\begin{equation}
    \label{eq:2}
    R = x \bar{\omega}_1 \omega_0 + (\bar{x} \oplus C) \omega_1 \omega_0
\end{equation}

\noindent where $-x$ is represented by the two-complement operation $\bar{x} \oplus C$ that consist of inverting the bits of x and adding C a constant corresponding to $2^{-d}$ with $d$ the decimal part size of the vector. Here, $\bar{x}$ now represents the bit-wise not operation. Applying De Morgan’s law on eq.~(\ref{eq:2}) will lead to the following:

\begin{align}
\label{eq:3}
R &= x \bar{\omega}_1 \omega_0 + \bar{x} \omega_1 \omega_0 \oplus C \omega_1 \omega_0,\\
\label{eq:4}
R &= \omega_0 (x \oplus \omega_1) \oplus C \omega_1 \omega_0,\\
\label{eq:5}
R &= \omega_0 (x \oplus \omega_1),
\end{align}

\noindent In eq.~(\ref{eq:4}), we consider $C=0$ which introduces an absolute error of $-2^{-d}$ when the MSB and the LSB of $\omega$ are equal to 0. The aim of such approximation is that eq.~$(\ref{eq:5})$ can be implemented by a single LUT3 component. Consequently, achieving 10,000 operations require $k \times 10^{4}$ LUT3 where $k$ represents the precision of $x$ and is chosen to lower the required resources and error. 

\subsubsection{First adder stage}

\begin{figure}[ht]
    \centering
    \includegraphics[width=1.0\textwidth]{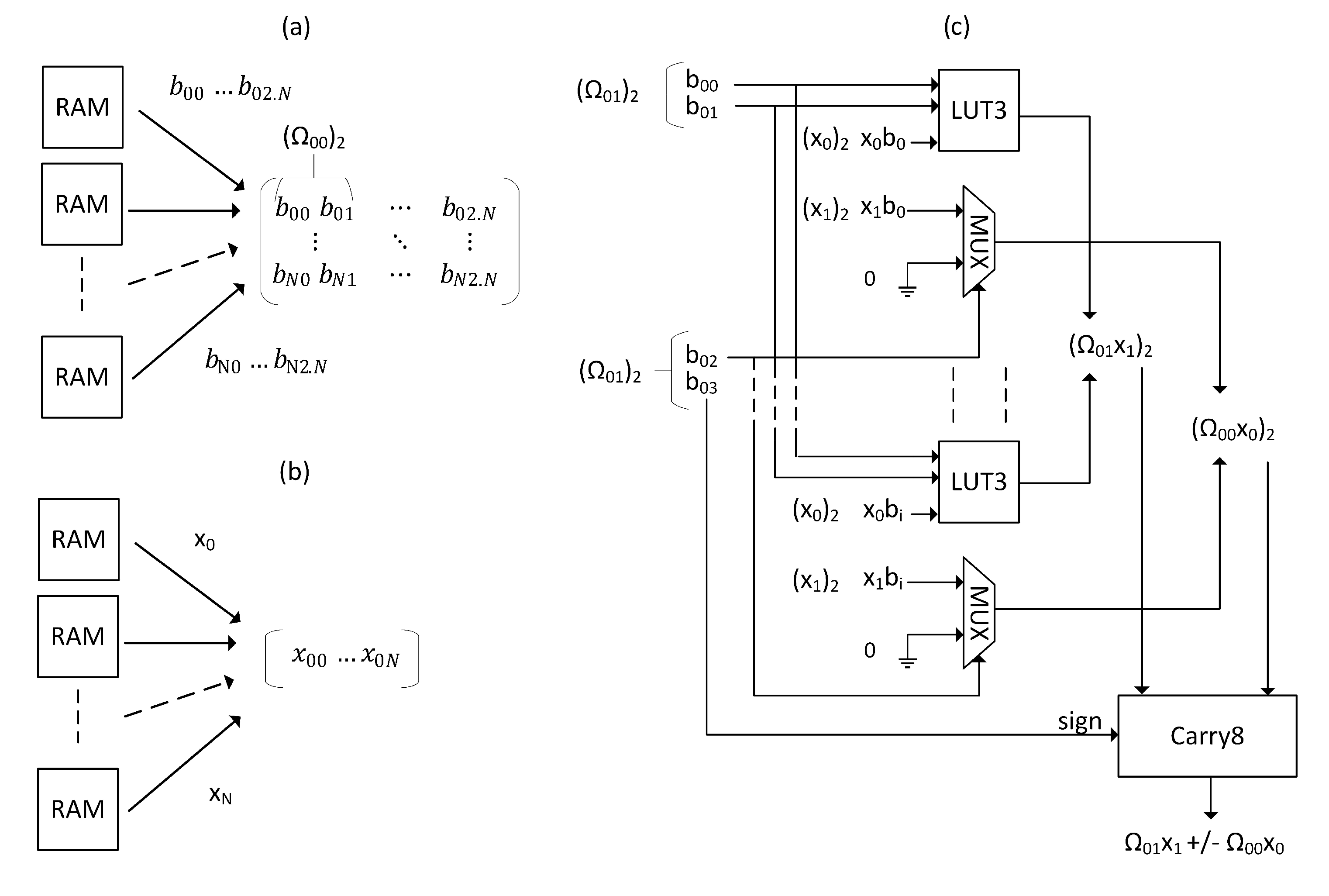}
    \caption{Representation of the high speed and multiple memory access apply to a circuit used for the dot product. (a) 100 RAM are instantiated and can be accessed at the same time. Each RAM corresponds to a row of the matrix $\Omega$. Each element $\omega_{ij}$ of $\Omega$ is encoded on a two bits vector. (b) As in (a), 100 RAM can be accessed at the same time to compose the vector x. (c) The $\Omega_{i,2j}$  and $x_{2i}$ values are injected into a LUT3 to compute the eq.~(\ref{eq:5}). The $x_{2i+1}$ values are injected into a multiplexer (MUX) whose selector is controlled by the LSB of $\Omega_{i,2j+1}$. The output of the LUT3 and the MUX are propagated through the CARRY8 component that add or subtract according to the value of $\Omega_{i,2j+1}$  MSB.}
    \label{fig:figure5}
\end{figure}

The circuit shown in Fig.~\ref{fig:figure5} represents the operations of multiplication used in the dot product based on eq.~(\ref{eq:5}). To accelerate the computation time, multiple access memory is utilized: Fig.~\ref{fig:figure5} (a) and (b) show the implementation of multiple block RAM (BRAM) that output 100 rows of 100 values at the same time. Eq.~(\ref{eq:5}) is implemented in LUT3 as described in the circuit of the Fig.~\ref{fig:figure5} (c) which compute the addition of two elements of the vector $\{\omega_{ij} x_i\}_{i}$. Using LUT3 for the elements at index $2i$, with i the index of the generated vectors beginning at 0, and multiplexers for the elements at index $2i+1$ ensures the use of lower resources and the optimal use of the configurable logic block (CLB). 

\subsubsection{DSP tree}
For block matrices and vectors of size $u$ with $u=100$, the dot product requires to perform 10,000 multiplications and 8,100 additions. The circuit of the Fig.~\ref{fig:figure5} (c) computes two multiplications and one addition. Thus, by reproducing this circuit 5,000 times, the FPGA computes 15,000 operations in a clock cycle and generates 100 vectors of 50 values that need to be added. This operation is done using an adder tree. A first stage of adder, that is not represented between the circuit of the Fig.~\ref{fig:figure5} (c), is realized using the CARRY8 component that reduces the 100 vectors of 100 elements to 100 vectors of 25 elements each. The remaining elements are then added using a DSP tree in adder mode. The advantage of using an adder tree of DSP is the low LUT number that is required, the optimal use of power consumption and the possibility to increase the frequency of the circuit. The Ultrascale architecture provided by the FPGA KU040 possesses a high number of DSP that can be cascaded allowing to reduce the routing circuit and the computation time. The DSP tree is repeated 100 times to compute at the same time all elements of the dot product resulting in the use of 1,200 DSP.

\subsubsection{Scalability of such architecture for bigger FPGA}

The number of clock cycles required to perform the dot product is given as follows:

\begin{equation}
\label{eq:6}
C_t(u,N) = K+(N/u)^2
\end{equation}

\noindent where $N$ is the problem size, $u$ the divider that partitions the matrix (in the current implementation u=100) and $K$ the number of clock cycles required for the multiplications and additions. The adder tree composed of CARRY8 and DSP previously proposed is constrained by the following condition:

\begin{equation}
\frac{N}{2^{h_C}+5^{h_D}}=1,
\end{equation}

\noindent where $h_C$ represents the height of the CARRY8 tree and $h_D$ the height of the DSP tree. The CARRY8 requires only one clock cycle when two cascaded DSP need 5 clock cycles. The height $K$ of the adder tree (in clock cycle) is then:

\begin{equation}
K=h_C+5h_D.
\end{equation}

\noindent We can fix $h_C$ as a constant according to the proposed FPGA circuit and find $h_D$ as follow:

\begin{align}
N&=2^{h_C}+5^{h_D},\\
N-2^{h_C}&=5^{h_D},\nonumber \\
\text{log}(N-2^{h_C})&=h_D \text{log}(5),\nonumber \\
h_D &= \frac{\text{log}(N-2^{h_C})}{\text{log}(5)}.
\end{align}

\noindent Then the height $K$ is equal to:

\begin{equation}
K=h_C+5\frac{\text{log}(N-2^{h_C})}{\text{log}(5)}.
\end{equation}

The adder tree increases logarithmically if we assume that an infinite amount of resources is available. Also, we show here that the design can be significantly improved if $u$ become larger with more available resources.

\subsection{Ising energy circuit}
The Ising energy is computed at the same time as the main dot product of $x_{i}$ by $\omega_{ij}$ because it shares the same output from the RAM that store the $\omega_{ij}$. As for the dot product, the Ising energy has been computed using logic equations to fit in a minimal number of LUT. The logic equation for the multiplication of $\sigma$ by the matrix $\Omega$ can be described as follows:

\begin{align}
	\label{eq:8}
	S_1 &= (\omega_{1} \oplus \sigma_j) \omega_{0},\\
	\label{eq:9}
	S_0 &= w_{0},
\end{align}

\noindent where $\sigma_i$ is the sign of $x_{i}$ and S is a 2 bits vector representing the multiplication of $\sigma_i$ at index $i$ by $\omega$ ( representing $\omega_{ij}$) which is also represented by 2 bits whose index is either 0 (LSB) or 1 (MSB). \newline

A circuit is also used to compute the Ising energy. Note that the Ising energy of a matrix M divided into matrices $m_{ij}$, where i and j are the indexes of the partitioned M, is equal to the sum of the energies of $m_{ij}$. Thus, the output of the pipelined Ising energy circuit is added with itself.

\subsection{Non-linear term}

\begin{figure}[ht]
    \centering
    \includegraphics[width=0.75\textwidth]{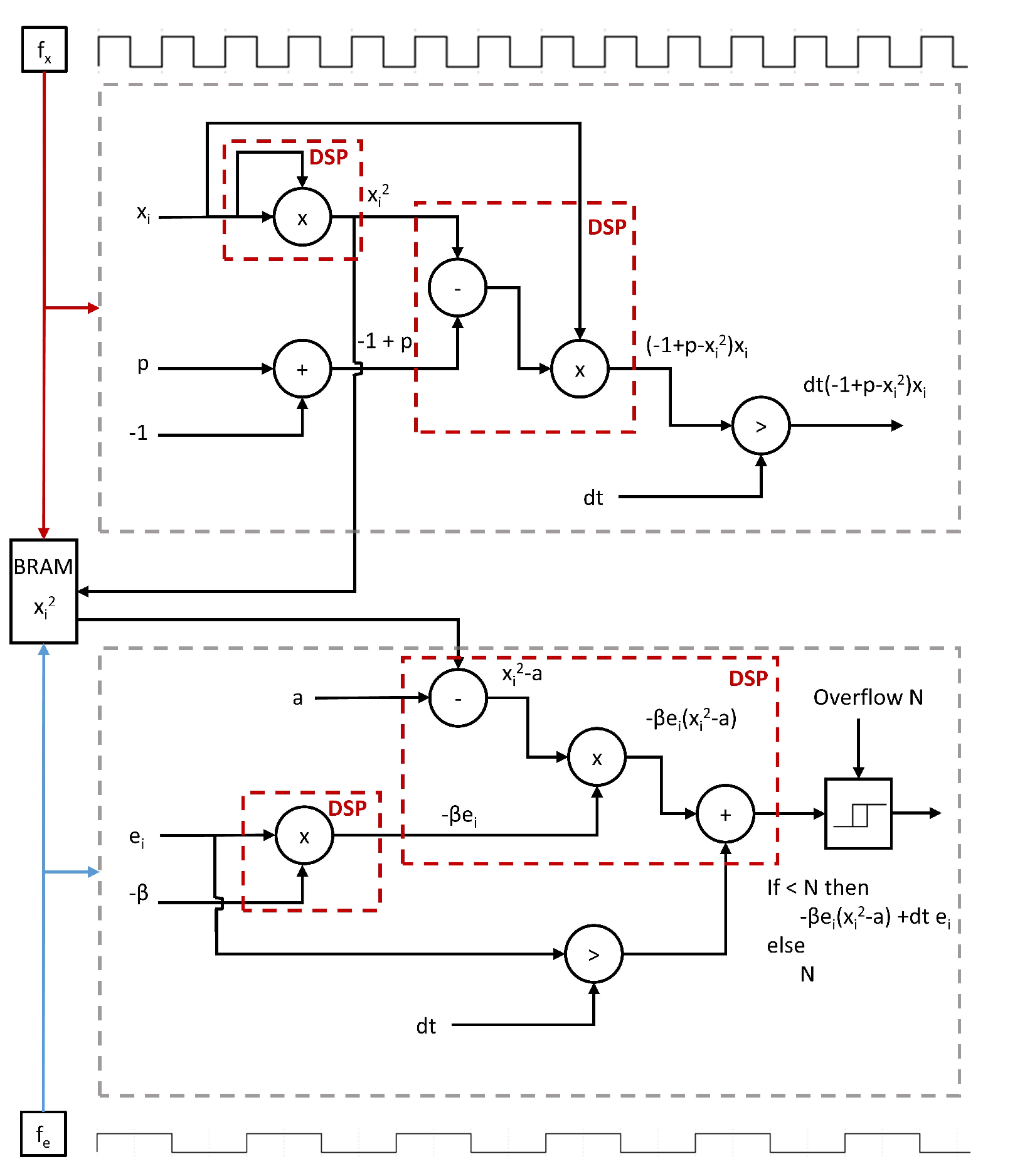}
    \caption{Circuits of the non-linear terms $x_i$ and $e_i$. The two terms use different clocks and share their results through a dual port block RAM allowing to read and write at different speed. Both circuit use DSP for high speed computation and are generated one hundred times to perform parallel computation. }
    \label{fig:figure8}
\end{figure}

To optimize the use of the electrical power, $x_i$ has been designed to use the highest frequency $f_x$ and the error correction use and intermediate frequency $f_e$. Both are computed several times during the operation of the dot product to reduce the computation time. Fig.~\ref{fig:figure8} shows the circuit use to compute one element of the vectors $x_i$ and $e_i$ that are reproduced 100 times. The circuits use pipelined DSP to compute additions, subtractions and multiplications. A shift register is used to multiply by the $dt$ of Euler approximation. To reduce the number of DSP into the design, $x_{i}^{2}$ is shared between $x_i$ and $e_i$ through a true dual port RAM that can be used with two different clocks to synchronize the two circuits. The RAM is controlled by an external FSM in the control module of Fig.~\ref{fig:figure1}.

\subsection{Power consumption}

Energy consumption of the circuit is determined by the number of logic transitions (from 0 to 1 or 1 to 0) and the frequency with P = $\langle s \rangle CV^2F$ where $P$ is the power dissipated by a transistor based circuit, $C$ the switching capacitance, $V$ the voltage and $F$ the frequency, and $\langle s \rangle$ the average number of switch per clock cycle. Voltage and capacitance are FPGA dependent since it is already manufactured. The digital circuit are at the highest power consumption when the components are enabled and when the clock signal drives the Configurable Logic Block (CLB) or the DSP. Thus, Enable and disable the block RAM is efficient to reduce the consumption however, clock gating shows better performances in this implementation. The methods have been extended on the available FF of the Xilinx FPGA and DSP which possess clock enable gate. However, when the design becomes large, high number of signal propagating enable towards CLB or DSP increase fanout and routing complexity. This can be solved by using BUFGCE component that are available in the FPGA and able to enable or disable the clock. Thus, when a given circuit is not needed, the clock can be disabled which will reduce significantly the power consumption.

The KU040 boards use Infineon regulators IR38060 incorporated voltage and current sensors. These sensors can be access either from the FPGA or from external bus with the PMBUS protocol which is based on i2c. We used here an Arduino to communicate with the regulators and record power data.

The power has been measured for different problem sizes and does not exceed 5W. Experiments show that the most critical operation for energy efficiency and computation time is the dot product which dissipates most of the FPGA power and needs more clock cycles to operate.\newline

\clearpage

\subsection{Parameter values used for solving SK spin glass instances}

The parameter values used for solving SK spin glass instances are shown in Tab. \ref{tab:param}.

\begin{table}[ht]
\centering
  \begin{tabular}{ | l | c | r |}
    \hline
    Symbol & meaning & value \\ \hline \hline
    $\beta$ & coupling strength & $0.25$  \\ 
		$\alpha$ & target amplitude baseline & $3.0$  \\
		$p$ & linear gain & $0.8-(\frac{N}{220})^{-2}$  \\
		$\rho$ & amplitude and gain variation & 3  \\
		$\delta$ & sensitivity to energy variations & 10  \\
		$\gamma$ & rate of increase of $\xi$ & 0.00011  \\
		$\tau$ & max. time w/o energy change & 1000 \\
		$n_x$ & number of iterations for nonlinear terms & 6 \\
		$n_e$ & number of iterations for error terms & 3 \\
		$dt_x$ & normalized time-step of nonlinear terms & $2^{-6}$ \\
		$dt_e$ & normalized time-step of error terms & $2^{-4}$ \\
		\hline
  \end{tabular}
\caption{\label{tab:param} Parameters used for solving SK problem instances.}
\end{table}

\clearpage

\begin{figure}[ht]
    \centering
    \includegraphics[width=1.0\textwidth]{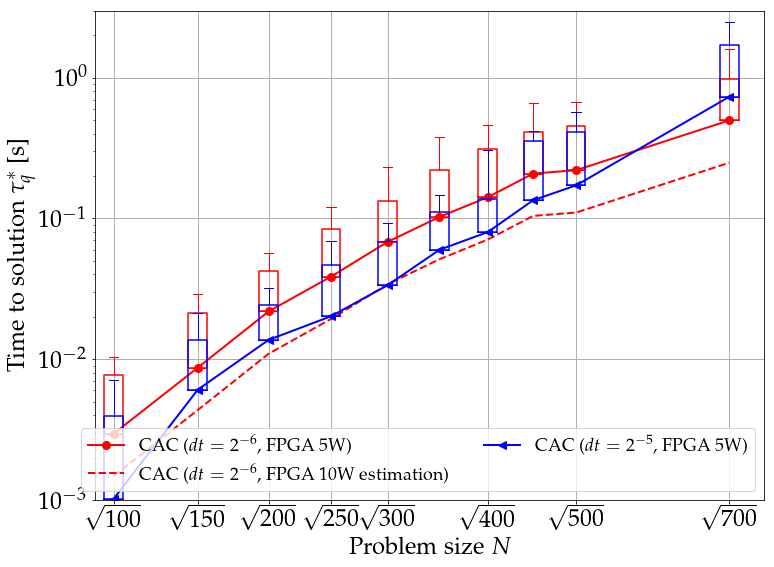}
    \caption{Lower, higher, and upper whisker of boxes show the $50^{\rm th}$, $80^{\rm th}$, and $90^{\rm th}$ percentiles of the real time to solution distribution in seconds for the FPGA implementation of CAC with a maximum of 5W power consumption with $dt_x = 2^{-6}$ and $dt_x = 2^{-5}$, and estimation of FPGA implementation of CAC with a maximum of 10W power consumption with $dt_x = 2^{-6}$.}
    \label{fig:figure10b}
\end{figure}

It is important to note that increasing the Euler step $dt_x$ does not always decrease the time to solution for all problem sizes; in particular, the time to solution of large problem sizes ($N=700$) is not significantly reduced when using $dt_x= 2^{-5}$ instead of $dt_x= 2^{-6}$ (see Fig.~\ref{fig:figure10b}). This is likely because the Euler approximation for larger problem sizes are prone to numerical instability for larger integration time steps.

\section{Benchmark on GSET}

Parameter values used for solving instances of the GSET are shown in Tab. \ref{tab:param2}.

\begin{table}[ht]
\centering
  \begin{tabular}{ | l | c | r |}
    \hline
    Symbol & meaning & value \\ \hline \hline
    $\beta$ & coupling strength & $\frac{3}{d_0}$  \\ 
		$\alpha$ & target amplitude baseline & $3.0$  \\
		$p$ & linear gain & $1-400 d_1^{-2.5}$  \\
		$\rho$ & amplitude and gain variation & $1.0$  \\
		$\delta$ & sensitivity to energy variations & $\frac{2.6}{N}$  \\
		$\gamma$ & rate of increase of $\beta$ & $\frac{2}{N}$  \\
		$\tau$ & max. time w/o energy change & $9N$ \\
		$n_x$ & number of iterations for nonlinear terms & 6 \\
		$n_e$ & number of iterations for error terms & 4 \\
		$dt_x$ & normalized time-step of nonlinear terms & $2^{-6}$ \\
		$dt_e$ & normalized time-step of error terms & $2^{-4}$ \\
		\hline
  \end{tabular}
\caption{\label{tab:param2} Parameters used for solving GSET problem instances.}
\end{table}

\noindent where $d_1$ is a function of the maximum degree given as $d_1 = \text{max}\{d_0,10\}$ and $d_0 = \text{mean}_i \{ \sum_j |\omega_{ij}| \}$.

\small

\clearpage

\section{Other algorithms}

\subsection{Details of NMFA simulation}

Noisy mean-field annealing can be simplified to the following discrete system\cite{King2018}:


\begin{align}
y_i(n+1) = (1-\alpha) y_i(n) + \alpha \text{tanh}[\frac{1}{\sigma_{\omega} T(t)}(\sum_j \omega_{ij} y_j(n)) + \sigma_r r_i],
\label{eq:NMFA}
\end{align}

\noindent with $\sigma_{\omega} = \sqrt{\sum_j J_{ij}^2}$. When the noise is not taken into account (i.e., $r_i = 0$), the steady state of eq.~(\ref{eq:NMFA}) is given as follows:

\begin{align}
y_i^* = \text{tanh}[\frac{1}{\sigma_{\omega} T(t)}(\sum_j \omega_{ij} y_j(n)) ],
\label{eq:NMFAsteady}
\end{align}

\noindent Note that the solutions of the eq.~(\ref{eq:NMFAsteady}) are the same as those of the TAP naive mean-field equations (see \cite{Bilbro1989}). Moreover, they are the same as the steady state of eq.~(1) when considering the change of variable $y_i = g(x_i)$ with $g(x) = \text{tanh}(x)$ and $\beta_i(t) = \frac{1}{T(t)}$. In fact, it can be shown that the two systems have the same set of attractors\cite{Pineda1988}.

The default parameters used in the numerical simulations are given as follows\cite{King2018}: $\alpha=0.15$ and $\sigma_r=0.15$.

\noindent Moreover, the temperature $T(t)$ is decreased with time according to an annealing schedule.

\begin{figure}[!b]
\begin{center}
\includegraphics[width=1.00\columnwidth]{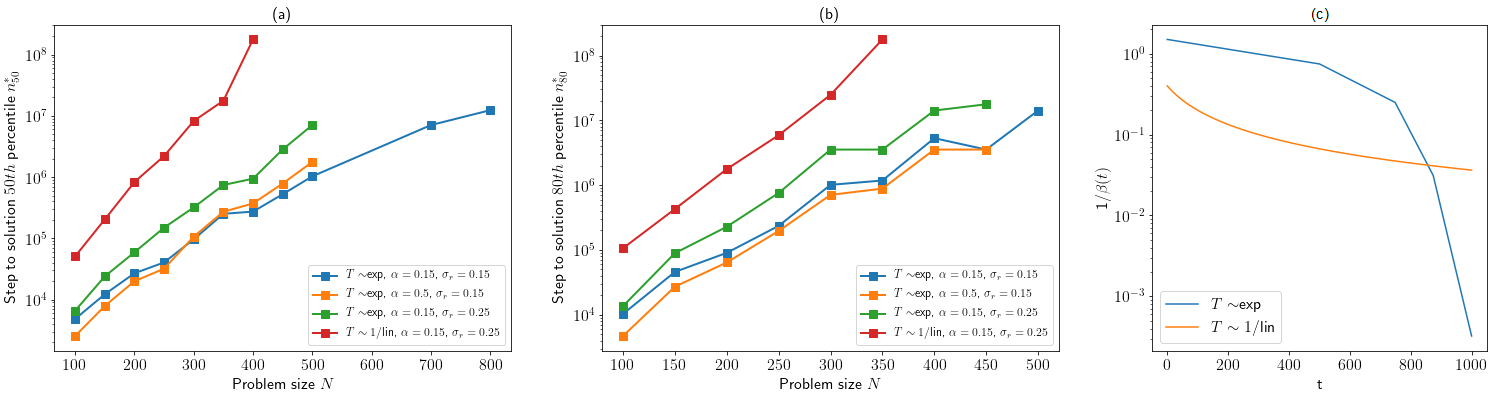}
\caption{\label{fig:suppNMFA} $50^{\rm th}$ (a) and $80^{\rm th}$ (b) percentiles of the step to solution distribution vs. the problems size $N$ of bimodal Sherrington-Kirkpatrick spin glass problems. (c) Exponential and inverse linear scaling of $T(t) = \frac{1}{\beta(t)}$ for $T=1000$.}
\end{center}
\end{figure}

The eq.~(\ref{eq:NMFA}) is simulated on a GPU using CUDA code provided in \cite{King2018}. Various parameters of the temperature scheduling $T(t)$ and parameters $\alpha$ and $\sigma_r$ have been tried in order to maximize the performance of this algorithm in finding the ground state of SK spin glass problems (see Figs. \ref{fig:suppNMFA}).

\clearpage

\subsection{Details of CIM simulation (simCIM)}

The physical model of the measurement feedback coherent Ising machine developed in \cite{Mcmahon2016} can be simplified as follows:
	
\begin{align}
x_i(n+1) =  A G(x_i(n)) + r_1 + \sqrt{\xi_0} \Theta(B \sum_j \omega_{ij} G(x_i(n)) + r_3))
\label{eq:simCIM}
\end{align}

\noindent where $\Theta(x) = R(-F x; x_{\text{max}})$ and $R(x;y)$ is the saturation function defined as $R(x;y)=x$ if $|x|<y$ and $R(x;y)=x_{\text{max}}$ otherwise. If we assume, for simplicity, that the saturation function $\Theta(x)$ is simply linear with $\Theta(x) = F_t x$, then eq.~(\ref{eq:simCIM}) can be written under the form of eq.~(1) by using the following: $f(x_i) = A G(x_i)$, $g(\beta x_i) = G(x_i)$, $\beta_i(t) = F(t) B \sqrt{\xi_0}$, and $r_1 + \sqrt{\xi_0} + r_3 = \sigma \eta_i$.

Eq.~(\ref{eq:simCIM}) is simulated using a GPU implementation in order to approximate accurately the success probability when it is small.

\bibliographystyle{naturemag}

\selectlanguage{english}
\FloatBarrier
\end{document}